# Interpretable Droplet Digital PCR Assay for Trustworthy Molecular Diagnostics

Yuanyuan Wei[1,2,†,*], Yucheng Wu[3,4,†], Fuyang Qu[1], Yao Mu[5], Yi-Ping Ho[1], Ho-Pui Ho[1], Wu Yuan[1,*], Mingkun Xu[3,6,*]

[1] Department of Biomedical Engineering, The Chinese University of Hong Kong, Shatin, Hong Kong SAR, 999077, China. E-mail: wyuan@cuhk.edu.hk.

[2] Department of Neurology, David Geffen School of Medicine, University of California, Los Angeles, California, 90095, USA. E-mail: yuanyuanwei@mednet.ucla.edu.

[3] Guangdong Institute of Intelligence Science and Technology, Hengqin, Zhuhai, 519031, China. E-mail: xumingkun@gdiist.cn.

[4] School of Computer Science and Technology, Hainan University, Haikou, 570228, China.

[5] Department of Computer Science, The University of Hong Kong, Hong Kong SAR, 999077, China.

[6] Center for Brain-Inspired Computing Research (CBICR), Department of Precision Instrument, Tsinghua University, Beijing, 100084, China.

[†] denotes equal contributions.

[*] Correspondence: xumingkun@gdiist.cn; wyuan@cuhk.edu.hk; yuanyuanwei@mednet.ucla.edu.

**Abstract**

Accurate molecular quantification is essential for advancing research and diagnostics in fields such as infectious diseases, cancer biology, and genetic disorders. Droplet digital PCR (ddPCR) has emerged as a gold standard for achieving absolute quantification. While computational ddPCR technologies have advanced significantly, achieving automatic interpretation and consistent adaptability across diverse operational environments remains a challenge. To address these limitations, we introduce the intelligent interpretable droplet digital PCR (I2ddPCR) assay, a comprehensive framework integrating front-end predictive models (for droplet segmentation and classification) with GPT-4o multimodal large language model (MLLM, for context-aware explanations and recommendations) to automate and enhance ddPCR image analysis. This approach surpasses the state-of-the-art models, affording 99.05% accuracy in processing complex ddPCR images containing over 300 droplets per image with varying signal-to-noise ratios (SNRs). By combining specialized neural networks and large language models, the I2ddPCR assay offers a robust and adaptable solution for absolute molecular quantification, achieving a sensitivity capable of detecting low-abundance targets as low as 90.32 copies/μL. Furthermore, it improves model's transparency through detailed explanation and troubleshooting guidance, empowering users to make informed decisions. This innovative framework has the potential to benefit molecular diagnostics, disease research, and clinical applications, especially in resource-constrained settings.

1. **Introduction**

The quantification of nucleic acids is fundamental to molecular diagnostics, especially in the detection and management of infectious diseases such as malaria[1], dengue[2], tuberculosis[3], and monkeypox[4]. Droplet digital PCR (ddPCR) is a highly sensitive and precise method that partitions PCR reactions into thousands of picolitre droplets to detect low-abundance targets[5,6]. It offers 2,000 times higher sensitivity compared to qPCR, enabling the detection of 1 mutant in 200,000 background wild-type genes without the need for standard curves[7]. This method has been instrumental in addressing global health challenges, including the rapid and reliable diagnosis of gene mutations[8–10] and diseases such as COVID-19[11,12], HIV[13], influenza[14], circulating tumor DNA[15–17].

In ddPCR, the copy number of target molecules is calculated by the fraction of the reaction units containing the target molecules, according to the Poisson distribution. Traditional ddPCR platforms face limitations such as high costs, operational complexity, and the need for manual data analysis, which can hinder widespread adoption and reproducibility[18,19]. Commercialized ddPCR platforms such as Bio-Rad's QX200[20] and RainDance RainDrop™[21,22], while precise in detecting targets as low as 0.1%[23], are expensive ( > $18,000) and require specialized expertise for operation and data interpretation. High initial and recurring expenses may deter use, especially in budget-strapped labs[18,19]. While flow cytometry stands as a revered 'gold standard' for cell analysis and sorting, it is less practical for ddPCR read-out[24]. The nature of ddPCR droplets, which are often water-in-oil emulsions, can pose significant incompatibility for flow cytometers optimized for aqueous samples. Meanwhile, maintaining ddPCR droplet integrity during flow cytometric analysis is challenging since shear forces that can cause coalescence, breakage, or deformation, resulting in sample loss and challenges for downstream analysis[26]. Manual fluorescence image analysis, reliant on software such as Image J[25], Fiji[26], and CellProfiler[27], is labor-intensive and lacks scalability[28,29]. These manual or semi-automated data preprocessing steps, such as thresholding and noise reduction, are also labor-intensive and prone to errors, especially in low-quality images with weak signals[24]. Automated methods, though promising, struggle with variability in experimental conditions, noise, and artifacts, limiting their generalizability and reproducibility across different laboratories[30,31].

Advancements in artificial intelligence (AI), including both machine learning and deep learning, have shown promise in automating and enhancing ddPCR image analysis[32–35]. AI-based methods, such as Attention DeepLabV3+ combined with circle Hough transform, achieve over 97% accuracy in droplet detection and classification, even in low-quality ddPCR images[36]. Another study employed convolutional neural networks (CNNs), achieving a remarkable 99.71% accuracy in classifying positive droplets[37]. A few-shot training technique requiring less dataset was applied for higher effectiveness[38]. By employing the YOLOv3 model, an accuracy of 98.98% was obtained while reducing the labeling time by 70% compared to the Mask R-CNN and YOLO models. Moreover, the strong generalized capability of neural networks has been validated in various ddPCR scenarios. Employing YOLOv5 and an RPN (Region Proposal Networks) model, the Deep-qGFP model has achieved precise and automated quantification of green fluorescence profiler-labeled droplets with an accuracy higher than 96.23% and a swift detection speed of 2.5 seconds per image[39]. Moreover, its strong generalized capability has been validated in various ddPCR formats, including droplet-based, microwell-based, and agarose-based ddPCR datasets. Furthermore, to eliminate the requirement of hand-crafted data or 'ground truth' for training, which leads to exhaustive data collection and manual annotation, training-free models such as the zero-shot Segment Anything Model (SAM) have also been incorporated[40]. These state-of-the-art (SOTA) AI models have advanced ddPCR image analysis through high automation levels, broad applicability, increased accuracy, personalized diagnosis, processing efficiency, and real-time capability.

Despite the advancements in AI-driven ddPCR image analysis, several critical challenges remain unresolved, which hinder the widespread adoption of automated ddPCR systems. These challenges primarily stem from the high variability in experimental conditions, the limited generalizability of current models, and their lack of interpretability. To be more specific, differences in reagent compositions, droplet generation settings, PCR cycling parameters, and fluorescence imaging conditions lead to significant variability in droplet morphology and fluorescence patterns, making it difficult for models to generalize across diverse setups [41,42]. Most existing models are trained on specific datasets and may struggle to adapt to new experimental settings, reducing their reproducibility and comparability across different laboratories. Moreover, Current AI models, while highly accurate, often lack transparency in their decision-making processes, making it difficult for users to trust and interpret their outputs.

To address these challenges, we introduce the intelligent interpretable droplet digital PCR (I2ddPCR) assay, a first-of-its-kind framework that integrates multimodal large language model (MLLM) with ddPCR image analysis model for nucleic acid detection and analysis. The key component of I2ddPCR assay, i.e., MLLM-ddPCR model, utilizes SAM for droplet recognition, a fluorescence-intensity-responsive classifier for droplet categorization, and the GPT-4o for producing interpretable diagnostic reports. Our framework not only attains accurate nucleic acid quantification but also offers detailed explanations of its decision-making process, aiding user comprehension and enhancing result reliability. The MLLM-ddPCR model outperforms SOTA models such as CNN or YOLOv5 models, achieving 99.05% accuracy within 7.37 seconds for complex ddPCR images containing over 300 droplets of different signal-noise-ratios (SNRs). In the inference phase, the system offers result interpretations and troubleshooting guidance through text, elucidating its decision-making process. The I2ddPCR assay is potentially able to reduce the variability in clinical decisions and harmonizes the biologists' experience with machine assistance. The I2ddPCR assay has the potential to decrease variability in clinical decisions and harmonizes biologists' expertise with machine assistance.

## 2. Results

**Workflow and Experimental Setup**

**Figure 1** provides a detailed overview of the workflow. For the initial demonstration, we adopted Seahorse (*Hippocampus kuda*) genome extracts, specifically the cytochrome c oxidase subunit I (*COI*) region (206 bp amplicon size), as the target template DNA. Serial dilutions (0.4 pg, 4 pg, and 40 pg within a 20 μL PCR mixture) were prepared for droplet generation, as summarized in **Figure 1a** (reagent details can be found in **Materials and Methods**). For each template, ~20,000 monodispersed droplets were generated (56.90 ± 0.55 µm in diameter, equivalent to a volume of 96.47 pL) using a customized lab-on-a-chip system. The droplets were collected into PCR tubes and then amplified inside a PCR amplifier for 45 cycles to achieve sufficient amplification. After the thermocycling process, SYBR Green dye was added and ~2,000 droplets were placed into a PDMS chip for observation. Spatial-lapse imaging was performed using a high-resolution FITC (Fluorescein isothiocyanate) fluorescence microscope to capture both bright-field images and fluorescence images.

To ensure the diversity of the testing dataset and avoid data leakage, we prepared ddPCR images using 9 distinct biological samples with variations in template and primer sequence design (see **Materials and Methods** for details). Disposable microfluidic chips with the same geometry were applied to avoid cross-contamination. Droplet size and uniformity were adjusted by fine-tuning the flow speeds of water phase and oil phase. For each sample, consistent imaging conditions, including focusing, fluorescence intensity, and magnification, were maintained. To ensure interframe continuity and image representativeness, we employed random sampling and avoided droplets overlapping.

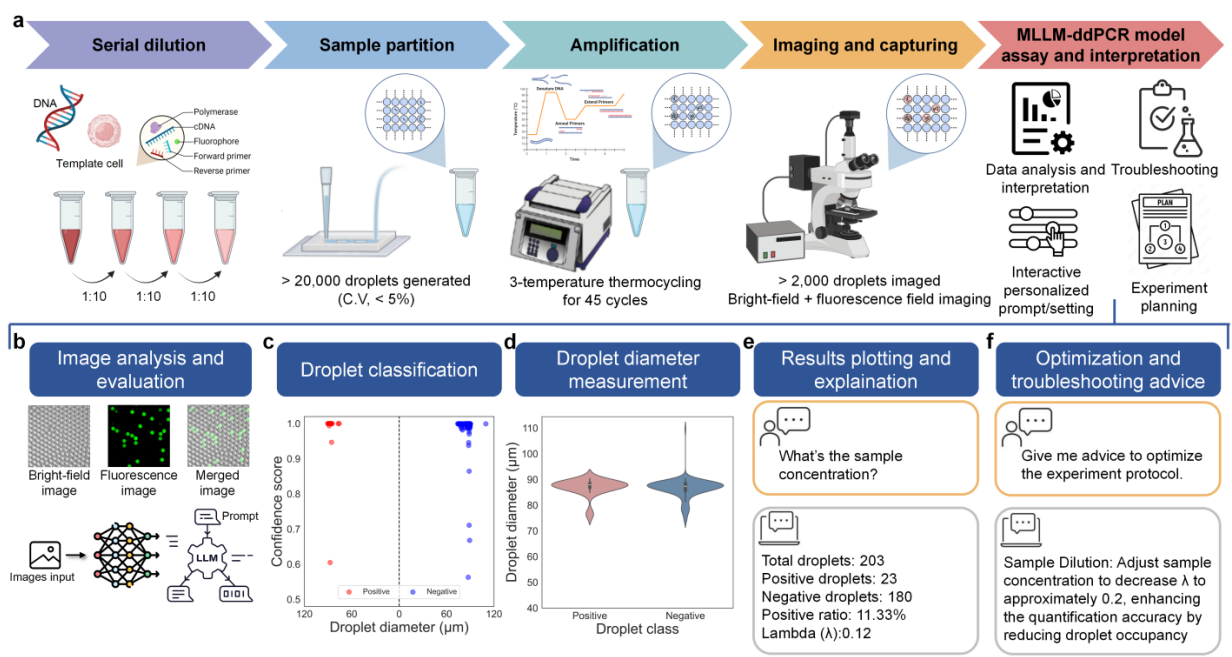

**Figure 1. I2ddPCR assay workflow. a.** Sample preparation workflow. The nucleic acid sample undergoes dilution, partitioning, amplification, and imaging for ddPCR result collection. The MLLM-ddPCR model then performs multiple functions including image analysis, context-aware explanation, and troubleshooting functions. **b–f.** MLLM-ddPCR model processing steps. **b.** Bright-field, fluorescence, and merged images are input via GUI for droplet segmentation and classification. **c.** The model converts the input images to droplet classification maps, including size distribution, and classification confidence. Positive and negative droplets are labeled in red and blue colors, respectively. **d.** The processed results are then used to reveal droplet diameter and droplet uniformity measurement. **e.** MLLM-ddPCR GUI automatically generates results (including droplet counts and calculated concentration factor $\lambda$) within 7.5 seconds and corresponding explanation within 30~40 seconds. **f.** MLLM-ddPCR model also automatically gives optimization and troubleshooting advice for aiding experiment design (result corresponding to **Supplementary Figure S7**) It also supports follow-up questions.

**Ground Truth Validation and Image Analysis**

To establish a reliable ground truth, the captured images were manually annotated by trained laboratory personnel with expertise in digital PCR and image analysis. These annotations were performed using *Roboflow*, designed for precise object segmentation and classification. The droplet diameter measurement was performed using the software *ImageJ* (National Institutes of Health and the Laboratory for Optical and Computational Instrumentation). To ensure the accuracy and consistency of the ground truth, the annotations were cross-checked by a separate group of experts who independently reviewed and verified the labels.

**Figure 1b-f** illustrate the MLLM-ddPCR model processing steps. The model integrates three main modules: the segmentation network (s-net), the classification network (c-net), and the MLLM. Upon inputting bright-field and fluorescence images (**Figure 1b**), the s-net segments droplets, while the c-net classifies them as positive (red) or negative (blue) based on fluorescence intensity (**Figure 1c**). Metrics include 'diameter/(μm)', and 'confidence score', with higher scores indicating more reliable classification. Simultaneously, droplet diameter was measured to calculate droplet volume and evaluate uniformity (**Figure 1d**). The MLLM generates context-aware explanation of the results, including droplet counts, size distribution, and concentration factor λ (**Figure 1e**, calculation details see **Materials and Methods**). Specifically, for high template concentrations ($\lambda > 1$), mathematical corrections are applied using $P_r(X = 2)$. The model also provides optimization and troubleshooting advice correspondingly (**Figure 1f**). The whole process is shown in a custom-developed graphical user interface (GUI, **Supplementary Figure 1**).

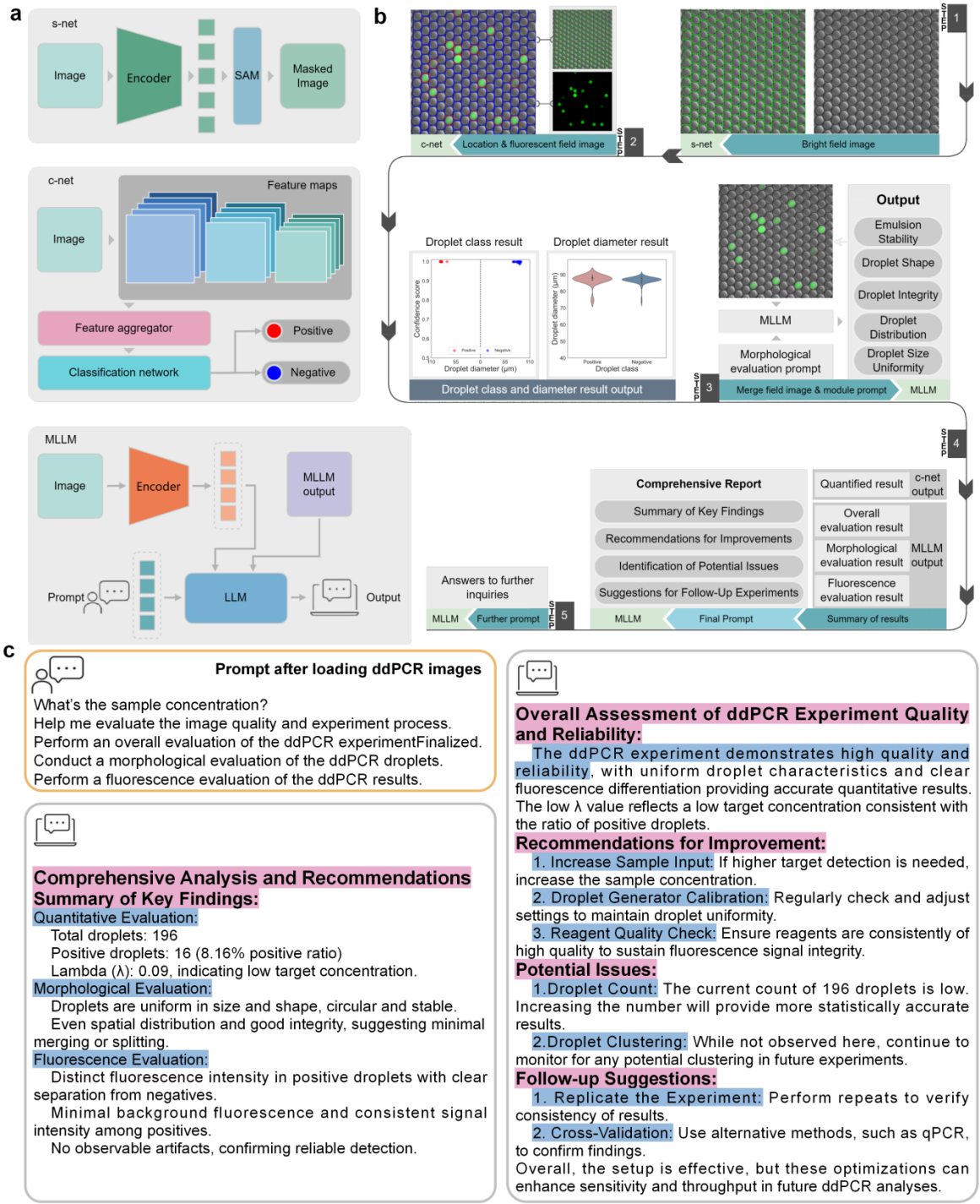

**Figure 2 MLLM-ddPCR model framework. a.** MLLM-ddPCR model structure includes three main neural networks: the segmentation network (s-net), the classification network (c-net), and the Multimodal Large Language Model (MLLM). **b.** ddPCR image analysis workflow: The bright-field image is segmented by s-net to detect droplet regions and measure droplet diameter. Each droplet in the fluorescence image is classified by c-net simultaneously for plotting classification results. The input images and user prompts are fed into MLLM to produce analytical results for each module. The results from the three modules and c-net are integrated and then input into MLLM to generate a final report. Asking follow-up questions is supported. **c.** Illustration of the detailed analysis and recommendations generated by the MLLM-ddPCR model highlights positive ratio evaluation, droplet uniformity, classification accuracy, and protocol optimization. The model recommends sample dilution, droplet stabilization, and PCR condition adjustments to enhance assay performance. It also identifies potential problems like DNA template sufficiency and primer dimer formation, suggesting targeted optimizations for improved ddPCR outcomes.

## Model Architecture and Representative Results

**Figure 2a** demonstrates the framework in detail. The s-net uses SAM for droplet detection and segmentation. SAM has been pre-trained on over 1 billion masks from 11 million images, thus requiring minimal retraining[43]. The c-net employs a fluorescence-intensity-responsive classifier for droplet classification. This classifier was trained on 2898 droplets (1176 positive droplets and 1722 negative droplets) and tested on 400 droplets (200 positives, 200 negatives) initially. this dataset augmented to 4,174 droplets (1,937 positive, 2,237 negative) for training and 604 droplets (302 positive, 302 negative) for testing subsequently to enhance capability of classifier. After the s-net and c-net, the MLLM provides results analysis and interacts with users by feeding prompts into GPT-4o. These networks are mutually conditioned and complementary for final diagnosis. Droplet candidates were manually confirmed using the GUI during training.

**Figure 2b** illustrates the workflow of MLLM-ddPCR model. The input bright field image is segmented by the s-net to detect droplet regions. After mask generation, the c-net maps the droplet positions onto fluorescence images and classifies droplets as positive or negative. This automated process yields droplet radius, fluorescence intensity, and classification data. The integration of segmentation and classification in a single pipeline ensures high accuracy and efficiency, addressing the challenges of droplet variability and noise in ddPCR images. MLLM-ddPCR then generates relevant plots and organizes quantitative analysis results for MLLM explanation. The explanation process is designed to provide transparent and actionable insights, overcoming the limitations of traditional ddPCR analysis methods or other SOTA AI models. It consists of two stages: Module Analysis and Second Diagnosis. In Module Analysis, ddPCR image analysis is divided into Overall, Morphological, and Fluorescence Evaluations, enhanced by in-context learning. This modular approach allows the model to focus on specific aspects of the image, ensuring comprehensive and accurate analysis. Comprehensive Analysis combines quantitative results and module outcomes into prompts for the MLLM, generating the final analysis report. This step leverages the GPT-4o's ability to explain complex data and provide human-like reasoning, enabling users to understand the results intuitively. Users can ask follow-up questions if needed, further enhancing the model's interactivity and usability.

**Figure 2c** shows the prompt for MLLM-ddPCR model as well as its representative results. The report covers five key aspects: Positive Ratio Evaluation, Droplet Uniformity Assessment, Droplet Classification Accuracy, Detailed Troubleshooting Advice, and Protocol Optimization. These outputs are designed to address common challenges in ddPCR experiments, such as variability in droplet size, fluorescence noise, and experimental setup inconsistencies. The troubleshooting section addresses potential issues such as DNA template sufficiency, PCR inhibition, and primer dimer formation, among others. This feature is particularly for users who may lack expertise in ddPCR optimization, as it provides actionable recommendations to improve experimental outcomes. The protocol optimization recommendations focus on adjusting reagent concentrations, reviewing primer design, and optimizing PCR cycling conditions. These suggestions are generated based on the model's analysis of the experimental data, ensuring that they are customized to the specific conditions of each experiment. The conclusion summarizes the current analysis, highlighting the need for sample dilution, droplet stabilization, and ongoing optimizations to enhance future ddPCR runs. To the best of our knowledge, it's the first time that an AI model has been applied for explaining and troubleshooting ddPCR images, demonstrating the potential of AI to revolutionize molecular diagnostics.

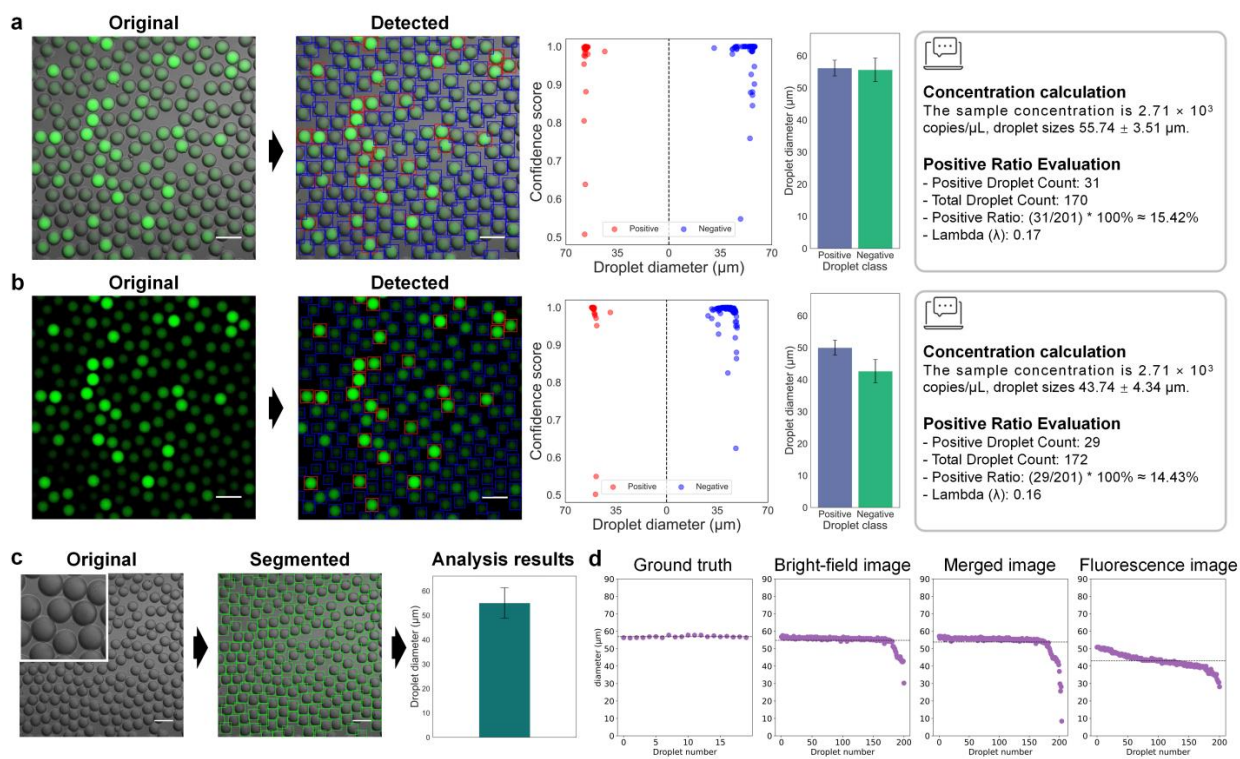

**Figure 3.** Performance of the I2ddPCR assay under varying imaging conditions. **a.** Merged images and **b.** FITC fluorescence images from benchtop ddPCR experiments are input for droplet diameter measurements and classification testing, respectively. Output includes labeled images, classification, size distribution stats, and explanation. **c.** A bright-field image of the same experiment, illustrating segmentation and droplet diameter measurement results. **d.** Bright-field images yield superior results due to reduced scattering effects. Scale bar: 100 μm.

**Robustness Across Imaging Modalities and Limit of Detection (LoD)**

Given the significant impact of varied experimental conditions and image quality on detection outcomes, we rigorously evaluated these effects on our model by determining the Limit of Detection (LoD) through adjustments in sample concentrations, imaging conditions, droplet count per image, and acceptable input image size. The LoD was determined to be 90.32 copies/µL to 2400.90 copies/µL (**Supplementary Figure S3**), demonstrating the system's sensitivity and reliability in detecting low-abundance targets. **Figure 3** shows the robustness of MLLM-ddPCR across bright-field, fluorescence, and merged images. Fluorescence and merged images both enable droplet classification and concentration inference (**Figure 3a** and **b**), but fluorescence images exhibit larger droplet diameter variability, leading to higher error rates. **Figure 3c** illustrates droplet segmentation and diameter measurement using a bright-field image. **Figure 3d** indicates that bright-field images yielded superior droplet diameter measurements (2.73 µm error) compared to fluorescence (22.95 µm error) and merged images (2.90 µm error), due to their absorption-based contrast.

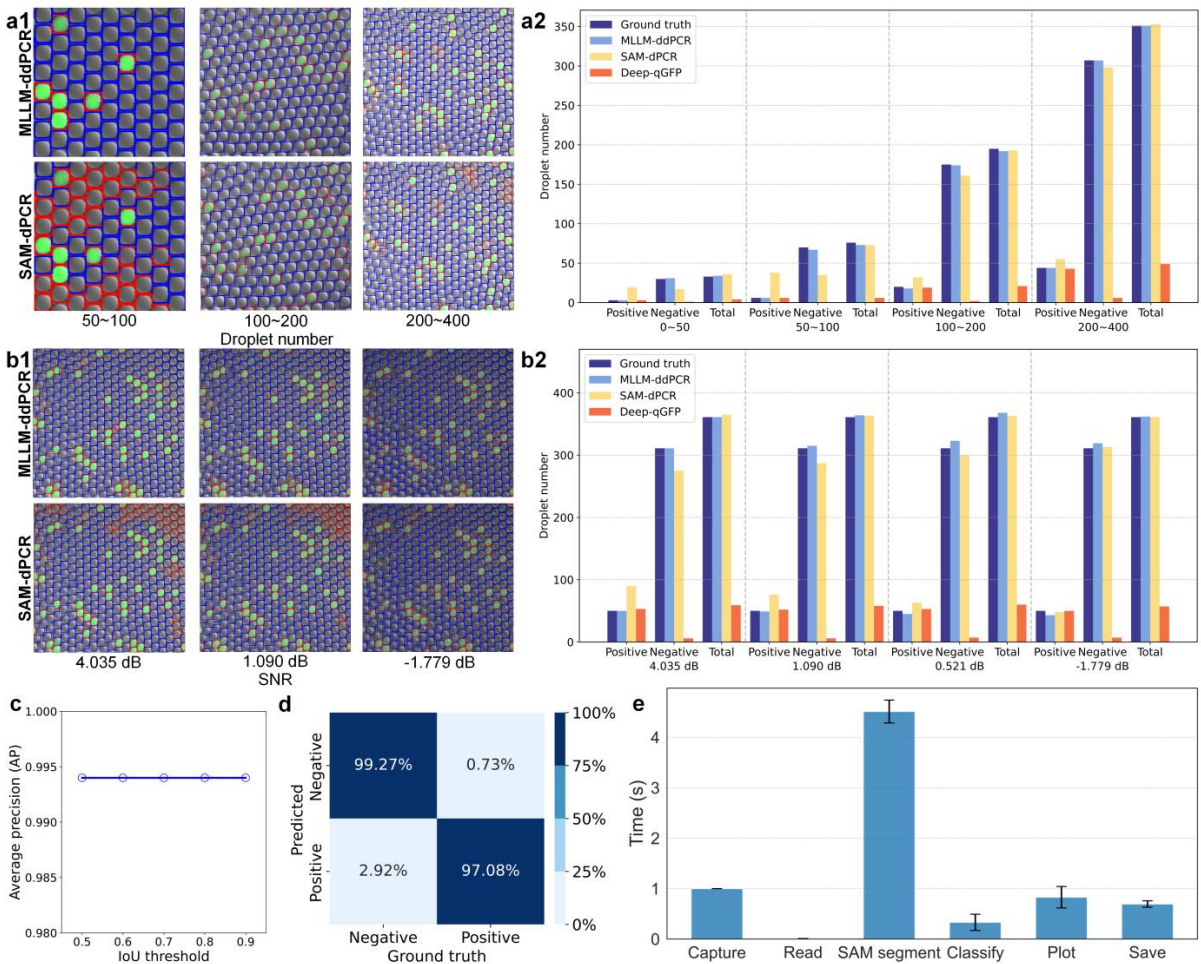

**Figure 4. Comparative performance of ddPCR image analysis models. a.** Comparative performance of models in different droplet counts per image. **a1.** Representative output labeled images by MLLM-ddPCR model, SAM-dPCR model, and Deep-qGFP model. **a2.** Comparison of droplet classification results. **b.** Comparative performance under different SNR levels, including **b1** representative output labeled images and **b2** droplet classification results. **c.** Average precision (AP) of MLLM-ddPCR model for different IoU thresholds. **d.** Visual representation of the confusion matrix for MLLM-ddPCR model. **e.** Time cost of MLLM-ddPCR model. Breakdown of the time cost for each step including Capturing, Reading, Segmenting, Classifying, Plotting, and Saving. The model processes hundreds of droplets per image within 7.37 seconds, excluding the capture time.

**Comparison with State-of-the-Art Models**

**Figure 4** compares MLLM-ddPCR with SOTA models (SAM-dPCR and Deep-qGFP). SAM-dPCR model is developed by incorporating zero-shot SAM and Otsu's method for auto-thresholding[40]. Deep-qGFP model was trained using a YOLO-v5m model and the Region Proposal Network (RPN) on over 200 manually labeled ddPCR datasets[44]. MLLM-ddPCR achieves 99.04% accuracy (**Supplementary Table 6**), outperforming Deep-qGFP (96.23%) and SAM-dPCR (97.10%). As shown in **Figure 4a**, MLLM-ddPCR accurately estimated droplet diameter with an average deviation of 1.473 pixels, surpassing 6.368 pixels for Deep-qGFP and 1.567 pixels for SAM-dPCR. We systematically altered image SNRs by introducing Gaussian and salt-and-pepper noise (**Figure 4b**). MLLM-ddPCR maintained higher precision (96.00% to 100.00%), compared to Deep-qGFP (83.01% to 85.41%) and SAM-dPCR (53.33% to 95.83%). Detailed experiment records for **Figure 4a** and **4b** can be found in **Supplementary Figure S12a** and **b**, respectively.

MLLM-ddPCR also demonstrated robustness against fluorescence noise, such as dust and air bubbles, with negligible impact on segregation accuracy. We intentionally introduced known amounts of dust (**Supplementary Figure S4b**) and air bubbles (**Supplementary Figure S4a1** and **S4a2**), finding no detectable impact on segregation accuracy. Neither contaminant was detected in the results, confirming their negligible effect. Moreover, we observed negligible spatial variations in background signal and found that the exposure could tolerate a certain degree of light transmission loss. For routine background signal evaluations, MLLM-ddPCR employs disposable microfluidic chips. As part of the LoD testing, we evaluated the performance of MLLM-ddPCR across various input sizes (1024×1024 to 96×96 pixels). While the system supports this wide range, resolutions below 256×256 pixels were found to compromise full image detection (**Supplementary Figure S5**). The optimal resolutions for accurate analysis were determined to be 1024×1024 or 512×512 pixels, corresponding to an active area of 1.264 × 1.264 mm² after edge exclusion.

MLLM-ddPCR achieved 97.93% mean average precision (mAP) at different thresholds of intersection over union (IoU), which quantifies the overlap between a predicted bounding box or segmentation mask and its corresponding ground truth annotation (**Figure 4c**). The fluorescence-intensity-responsive classifier's performance was tested on a dataset containing 302 positive and 302 negative droplet objects. It also reveals 97.35% total classification accuracy, with 97.08% and 99.27% accuracies for positive and negative classes, respectively (confusion matrix in **Figure 4d**). A detailed breakdown of MLLM-ddPCR's runtime is provided in **Figure 4e**. The process includes: (1) reading images from the fluorescence microscope's application programming interface (API) or software development kit (SDK) cable (0.006 seconds per image), (2) segmenting and classifying droplets (4.51 and 0.33 seconds per image, respectively), (3) plotting results (0.83 seconds per image), and (4) outputting results into a designated folder (0.04 seconds per image).

As a result, MLLM-ddPCR takes no more than 7.37 seconds for analyzing a ddPCR image, excluding the capture time.

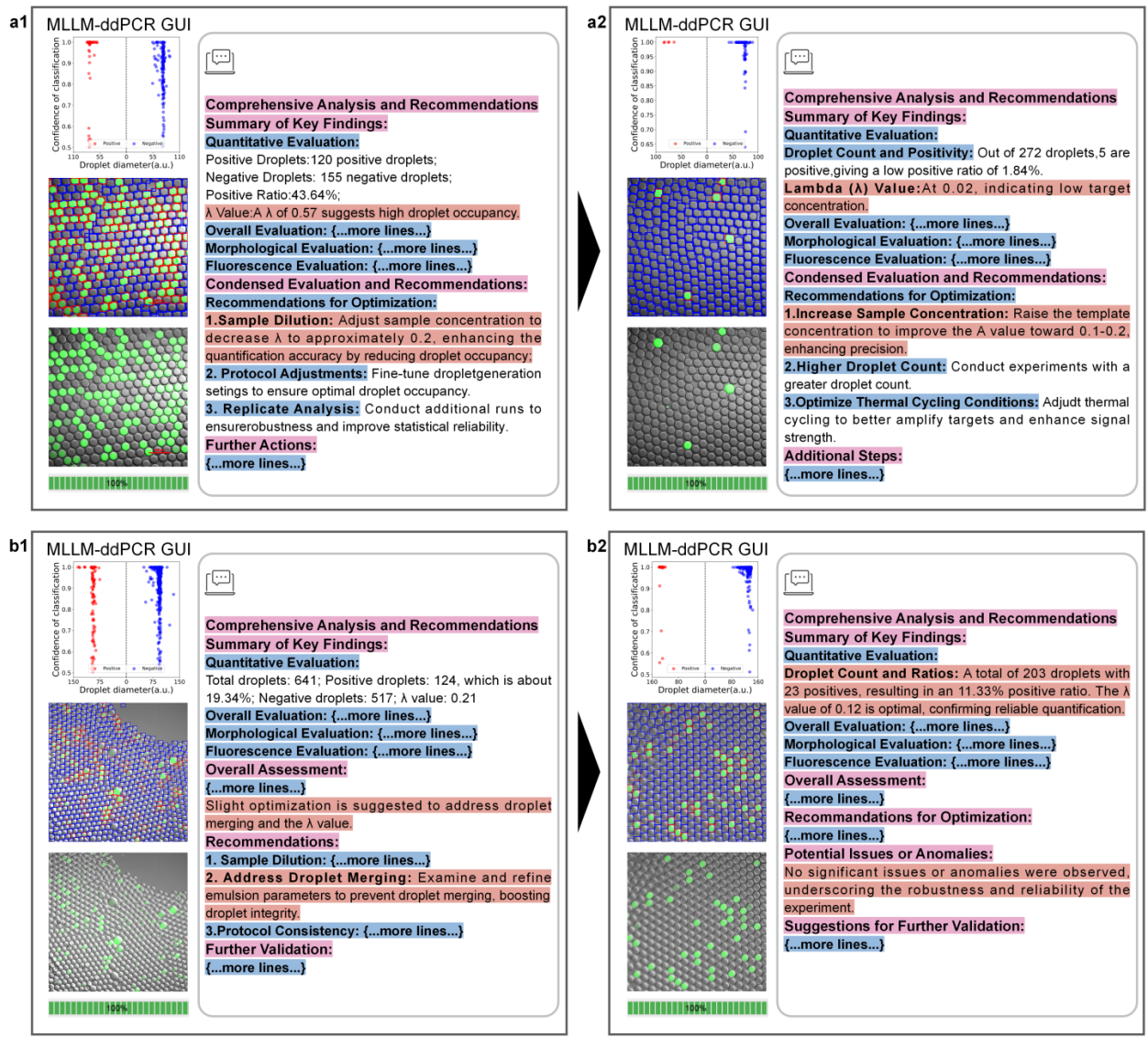

**Figure 5. Demonstration of I2ddPCR assay troubleshooting and iterative feedback mechanism. a.** Template concentration optimization. MLLM-ddPCR model identifies the initial ddPCR image. **a1**. as problematic. Following the model's advice to lower the template concentration from 2.00 pg/μL to 0.20 pg/μL, the updated ddPCR image. **a2**. was determined as low concentration. **b.** Droplet uniformity improvement. The MLLM-ddPCR model detects droplet size non-uniformity in the initial ddPCR image. **b1**. attributing it to droplet merging during generation and amplification processes. Following the recommendations for fine-tuning the droplet generation process and conducting replicate experiments for consistency, the explanation of the improved image. **b2**. confirms successful enhancement in droplet uniformity. Scale bar: 100 μm.

**Troubleshooting and Optimization of Experimental Conditions Enabled by MLLM-ddPCR**

**Figure 5** demonstrates the MLLM-ddPCR model's capability for troubleshooting and optimizing ddPCR experimental conditions. Initially, the model identifies the input ddPCR image (**Figure 5a1,** sample concentration of 2.00 pg/µL) as problematic since the positive ratio is 43.64 and $\lambda$ is 0.21. Under this condition, the positive ratio as well as $\lambda$ is too high for accurate quantification. The model's explanation suggests lowering the template concentration to address this issue. Following the model's advice, the experiment is adjusted to a lower template concentration of 0.2 pg/µL, resulting in a new ddPCR image **Figure 5a2**. The model's subsequent explanation of **Figure 5a2** confirms the successful correction, with a positive ratio of 1.84% and $\lambda$ of 0.02. Similarly, the initial ddPCR image **Figure 5b1** shows uniform droplets, which was attributed to droplet merging during generation and amplification processes in MLLM-ddPCR explanation. The model highlights the need for fine-tuning the droplet generation process, prompting an enhanced droplet uniformity ("droplet formation is consistent") in **Figure 5b2**. This iterative feedback mechanism demonstrates the model's ability to provide actionable insights, optimizing ddPCR protocols for accuracy and reliability, thereby advancing molecular diagnostics.

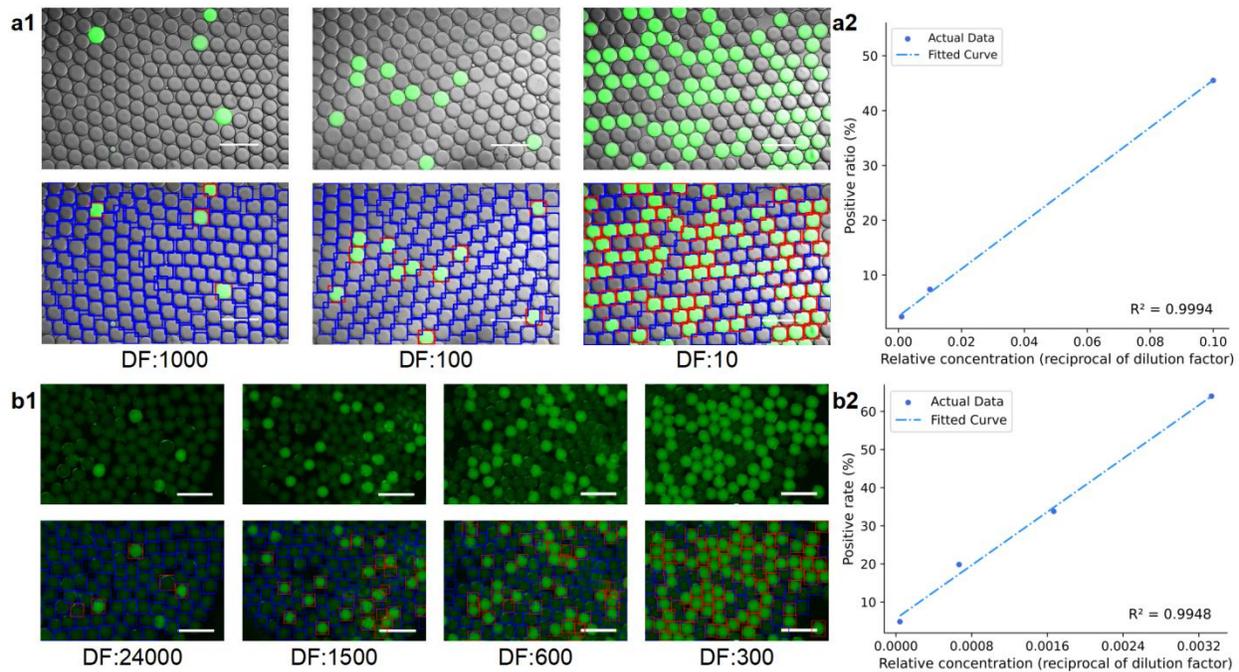

**Figure 6. Linearity and model generality in hydrogel-based dPCR image analysis. a.** Performance under varying ddPCR sample concentrations (0.4 pg, 4 pg, and 40 pg within a 20 µL PCR mixture). Strong linearity between known concentration and positive ratio is observed, with r² values reaching 0.9994. **b.** Hydrogel-based dPCR image analysis for different dilution factors

(DF) of harvested *S. Typhi* DNA. The linear regression between known concentration and positive ratio ($r^2 = 0.9948$) demonstrates the model's accuracy and robustness across a wide range of concentrations. Scale bar: 100 μm.

**Reliability and Generalizability Testing**

We tested the reliability of MLLM-ddPCR by yielding linear regressions between known concentration and positive ratios. **Figure 6a** shows the results of detecting templates of different concentrations. The analysis results yielded inferred concentrations ranging from 0.27 to 2.50 × $10^3$ copies μL$^{−1}$. A strong linearity of $r^2 = 0.9994$ was observed. Here $r^2$ represents the coefficient of determination, computed using the standard formula for linear regression (see **Methods**). Output plots for each input image can be found in **Supplementary Figure S13 and S14**). MLLM-ddPCR also exhibits robust generalization capabilities when directly applied to different dPCR scenarios without training. As demonstrated in **Figure 6b**, MLLM-ddPCR has been validated with hydrogel-based ddPCR images with serial dilutions of harvested *S. Typhi* DNA at 24000, 1500, 600, and 300 times. The linear regressions between known concentration and positive ratio yields an $r^2$ value of 0.9948. Specifically, the results for **Figure 6a** and **b** have been cross validated by comparing with manual counting results in **Supplementary Table 7**. These results demonstrate our model's generalizability and versatility, highlighting its potential to accelerate the identification of therapeutic targets and biomarkers.

## 3. Conclusions

In summary, we present the I2ddPCR assay facilitated by the MLLM-ddPCR model, which signifies a substantial progression in the realm of ddPCR data explanation. Our model's ability to segment and classify droplets, calculate DNA template concentrations using Poisson statistics, and evaluate image quality demonstrates its versatility and robustness. Integrated with MLLM, MLLM-ddPCR model offers an automated, comprehensive solution beyond mere quantification, providing insightful image analyses, experimental condition appraisals, and diagnostic guidance. MLLMs excel in processing vast amounts of information, identifying patterns and correlations that may elude human analysts. Their training on diverse datasets enables them to adapt to various ddPCR image characteristics and experimental conditions. Additionally, MLLMs provide context-aware explanations by integrating metadata about the experimental setup, such as reagents and target sequences. This enhances the accuracy of size distribution assessments and boosts the reliability of classification confidence metrics.

The comprehensive outputs, including detailed image explanations and experimental condition evaluations, empower researchers to make more informed decisions and optimize their ddPCR protocols. The benchmarking results against human expert analysis and state-of-the-art ddPCR image analysis models further validate MLLM-ddPCR's superior accuracy, efficiency, and reproducibility. Our demonstration shows wide-ranging DNA intercalating dye-labeling scenarios and extends across agarose-based dPCR experiments. Based on this, we believe that our method has strong generalizability for learning other droplet microfluidics-enabled molecular diagnostic patterns, such as digital loop-mediated isothermal amplification (dLAMP)[45–47], digital recombinase polymerase amplification (dRPA)[48–51], and droplet-based bacteria quantification[52,53]. As we continue to refine and expand the capabilities of our model, we anticipate that MLLM-ddPCR will become an indispensable tool for researchers seeking to harness the full potential of ddPCR technology. By integrating with other nucleic acid sensors[54], signal amplification biosensors[55], and point-of-care-testing (POCT) platforms such as smartphone-based portable ddPCR devices[56–58], I2ddPCR could form a cohesive platform that streamlines the entire research workflow. This deep-learning-enabled ddPCR analysis assay can be used with various molecular diagnostic assays and might help expedite research in disease detection and treatment development. Our method allows diagnostic consistency and cost-effective system deployment to meet clinically demanding needs, for example in a cloud-based host for providing objective second opinions and consensus in resource-limited areas. Furthermore, our method can be seamlessly integrated into an agentic AI framework [NMI paper] to enhance the autonomy of ddPCR analysis. This integration allows AI agents to access a variety of tools and collaborate in a multi-agent environment, fostering collective intelligence. For instance, within an agentic framework, our method can connect to laboratory information management systems (LIMS), facilitating real-time data tracking, analysis, and reporting. As we continue to refine and expand the capabilities of the MLLM-ddPCR model, we envision its integration into broader molecular diagnostic workflows, enabling seamless collaboration with other AI-driven tools and point-of-care platforms to revolutionize disease detection, treatment development, and personalized medicine.

## 4. Materials and Methods
### 4.1 Sample preparation for ddPCR experiments

PCR mixture for **Figure 3** experiments:

Components: 1X Platinum SuperFi II buffer, 0.2 mM dNTP, 1X Platinum SuperFi II polymerase, 0.5 µM forward and reverse primers, templates, 0.2% Tween 20, 0.2 mg/mL BSA (NEB, USA), and 0.4% PEG-8000.

PCR protocol: Initial denaturation at 98°C for 30 s, followed by 45 cycles of denaturation at 98°C for 10 s, annealing at 60°C for 10 s, and extension at 72°C for 15 s. Final extension at 72°C for 5 min and an indefinite hold at 12°C.

Templates and primers:

PCR template (5'→3'):

GTCTCGTGGAGCTCGACAGCATNNNNNNTGNNNNNNTGCCTACGACAAACAGACCTAAAATCGCTCATTGCATACTCTTCAATCAG

Forward primer: 5'-Acrydite-ACTAACAATAAGCTCUAUAGTCTCGTGGAGCTCGACAG-3'

Reverse primer: 5'-CTGATTGAAGAGTATGCAATGAG-3'

PCR mixture for **Figure 4**:

Components: 1X TaKaRa PrimeSTAR GXL buffer, 0.2 mM dNTP, 1X TaKaRa PrimeSTAR GXL polymerase, 0.2 µM forward and reverse primers, templates, 0.5% Tween 20, 0.1 mg/mL BSA (NEB, USA), and 0.5% PEG-8000.

PCR protocol: 25 cycles at 95°C for 10 s, 59°C for 15 s, and 68°C for 15 s, followed by a final extension at 68°C for 3 min and a hold at 25°C.

With similar experiment protocol, the PCR template sequence for **Figure 1** (5'→3') is:

GTCTCGTGGAGCTCGACAGNNNNNNNNNNNNNNNNNNAGTGCTACTCTCCTCGCTCC

Template concentration: 1.25 fM (0.25 uL, 0.1 pM template in 20 ul PCR mix).

Cycle number: 40.

The PCR template sequence for **Figure 2** (5'→3') is:

GTCTCGTGGAGCTCGACAGNNNNNNNNNNNNNNNNNNAGTGCTACTCTCCTCGCTCC

Template concentration: 1.25 fM (0.25 uL, 0.1 pM template in 20 ul PCR mix).

PCR mixture for **Figure 5a, 6a and S11** experiments:

The ddPCR mixture was prepared with 1×KAPA HiFi buffer and polymerase (KK2502, Roche, Switzerland), 0.3 mM dNTP, 0.3 µM reverse primers, 0.1% NP-40, 0.2% Tween 20, and 0.1 mg/mL BSA (NEB, USA), and varied amounts of templates.

Template Concentrations: 40 pg, 4 pg, and 0.4 pg per 20 µL PCR system.

Target: *Cytochrome c oxidase subunit I (COI)* gene from Seahorse (Hippocampus kuda), amplicon size of 206 bp. The design has been validated in our previous published work[59].

Primers:

Forward: 5'-TTTCTTCTCCTCCTTGCTTCCTCAG-3'

Reverse: 5'-GAAATTGATGGGGGTTTTATGTTG-3'

The PCR template for **Figure 5b** is:

_GTCTCGTGGAGCTCGACAGNNNNNNNNNNNNNNNNNNAGTGCTACTCTCCTCGCTCC

Template conccentration: 1.25 fM (0.25 uL, 0.1 pM template in 20 ul PCR mix).

Sequences of templates and primers for **Supplementary Table 7** are provided below.

**Table 1. Sequences of templates and primers in Supplementary Table 7.**

|  |  |  | Sequence (5'→3') |
|---|---|---|---|
| Barcode 1 | Template |  | GTCTCGTGGAGCTCGACAGNNNNNNNNNNNN TCGCTCATTGCATACTCTTCAATCAGC |
|  | Primer | Forward | GTCTCGTGGAGCTCGACAG |
|  |  | Reverse | GCTGATTGAAGAGTATGCAATG |
| Barcode 2 | Template |  | GTCTCGTGAGTCAGGACAGNNNNNNNNNNNN TCGCTCATTGCATACTCTTCAATCAGC |
|  | Primer | Forward | GTCTCGTGAGTCAGGACAG |
|  |  | Reverse | GCTGATTGAAGAGTATGCAATG |
| Barcode 3 | Template |  | GTCTCGTGACCTCGGACAGNNNNNNNNNNNN TCGCTCATTGCATACTCTTCAATCAGC |
|  | Primer | Forward | GTCTCGTGACCTCGGACAG |
|  |  | Reverse | GCTGATTGAAGAGTATGCAATG |
| Barcode 4 | Template |  | GTCTCGTGGACAGTGACAGNNNNNNNNNNNN TCGCTCATTGCATACTCTTCAATCAGC |
|  | Primer | Forward | GTCTCGTGGACAGTGACAG |
|  |  | Reverse | GCTGATTGAAGAGTATGCAATG |

We have ensured the stability of the samples throughout the ddPCR process and during storage at 4°C for 48 hours.

### 4.2 Microfluidic chip fabrication

A flow-focusing microfluidic chip with dimensions of 30.0 μm (width) and 38.5 μm (height) was designed and fabricated for droplet generation (**Supplementary Information Figure S15**). For chip design, we illustrated the patterns using AutoCAD software (Autodesk, San Rafael, USA). It was then outsourced to print on transparent films with a resolution of 25,000 DPI for mask preparation (MicroCAD Photo-Mask Ltd., Shenzhen, China). Then the chip was created using standard SU-8 photolithography (SU8-3050 photoresist from Kayaku Advanced Materials, Westborough, USA) and Polydimethylsiloxane (PDMS) replica molding processes. The patterns were transferred onto the 4" silicon wafers, followed by UV exposure, baking, and developer bath according to the manufacturer's specifications.

A mixture of PDMS prepolymers (base-to-curing weight ratio of 10:1, Dow Corning, MI, USA) was prepared and poured onto the SU-8 master molds. After curing at 80°C for 10 h, the crosslinked PDMS was then removed and cut to the desired shape. Inlets and outlets were punched using a 1.0 mm Miltex biopsy puncher (Integra Life Sciences, NJ, USA). Both the PDMS slabs and glass slides were treated with oxygen plasma for 1 min and bonded by brief baking at 110°C. The microfluidic devices were hydrophobized by baking at 55°C for 24 h.

### 4.3 Droplets production and imaging

In this work, we generated monodispersed droplet emulsions using a custom microfluidic setup. The droplet generation process was actuated by two syringe pumps (Legato 100, KD Scientific or

Ph.D. 2000, Harvard Apparatus, USA) at the inlets. The dispersed phase was ddPCR mixture and a commercial droplet generation oil (1864006, Bio-Rad Laboratories, Inc.) was used as the continuous phase. By using the generated microfluidic chip, droplets with a mean diameter of 56.90 ± 0.55 µm (0.096 nL) were produced with flow rates of around 10 and 14 µL/min for the water phase and oil phase, respectively. After collecting droplets in 0.2 mL PCR tubes, mineral oil was added to the top surface to prevent evaporation. The ddPCR reactions were conducted in a thermal cycler (T100, Bio-Rad, USA) with the following programming: Initial denaturation at 95°C for 3 min, followed by 45 cycles of denaturation at 98°C for 20 s, annealing at 61°C for 15 s, and extension at 72°C for 15 s. Final extension at 72°C for 1 min and an indefinite hold at 12°C. We have ensured the stability of the samples throughout the ddPCR process and during storage at 4°C for 48 hours.

After PCR amplification, the amplified droplets were collected and transferred into a custom PDMS chamber designed for observation. We used an inverted microscope (Nikon Eclipse Ti-U) with a camera (Nikon DS-Qi2) to capture both bright-field and fluorescence images (fluorescence excitation at 455 nm; emission at 495 nm).

**4.4 Development of MLLM-ddPCR model**

The MLLM-ddPCR algorithm integrates the SAM and GPT-4o architectures. The SAM model, based on a CNN, extracts robust features from 11 million images and 1.1 billion masks (SA-1B dataset, available at https://segment-anything.com). These masks, averaging 100 per image, are generated via automated segmentation and validated through human ratings. The GPT-4o model, utilizing an enhanced Transformer architecture, generates coherent text from a diverse dataset of trillions of words, trained using both supervised and reinforcement learning.(More information at https://openai.com/index/hello-gpt-4o/)The MLLM-ddPCR process is implemented using Pytorch on an NVIDIA Tesla V100-SXM2-16GB platform. Detailed instructions are available at https://github.com/WEI-yuanyuan/MLLM-ddPCR.

The template concentration is calculated as below:

The probability $\Pr(X = k)$ that a droplet will contain k copies of the target gene if the mean number of target copies per droplet is $\lambda$:

$$f(k, \lambda) = \Pr(X = k) = \frac{\lambda^k e^{-\lambda}}{k!} \qquad (1)$$

where

$k$ is the number of occurrences ($k$ can take values 0, 1, 2, ...).

$e$ is Euler's number ($e = 2.71828…$).

! is the factorial function.

Inputting $k=0$ gives the probability that a droplet will be empty:

$$\Pr(X = 0) = e^{-\lambda} \qquad (2)$$

For the number of droplets being large enough, the observed fraction of empty droplets (E) gives estimation of $\Pr(X = 0)$

$$E = e^{-\lambda} \tag{3}$$

At the same time, by definition of E,

$$E = \frac{N_{negative}}{N} \tag{4}$$

Solving (3) we get

$$\lambda = -\ln(E) \tag{5}$$

As $\lambda$ is the copies per droplet, concentration of copies per volume is

$$Concentration = \frac{\lambda}{V_{microreactor}} \tag{6}$$

Here we applied mathematical corrections for high template concentrations ($\lambda > 1$) by using a correction factor $P_r(X = 2)$:

$$\lambda' = \lambda + 2 \times P_r(X = 2) \tag{7}$$

$\lambda'$ is the corrected average number of target molecules per partition, considering partitions that may contain more than one template,

Thus for concentrations with $\lambda > 1$, their respective concentrations are corrected to:

$$Concentration = \frac{\lambda'}{V_{microreactor}} \tag{8}$$

To ensure transparency, we have included the detailed calculations and results of these probabilities for each experimental setup in **Supplementary Information Table 1.**

The accuracy of MLLM-ddPCR is calculated based on Equation 9. The calculation process can be found in **Supplementary Table 9**.

$$ACC = \frac{TP + TN}{TP + TN + FP + FN} \tag{9}$$

### 4.5 Graphical user interface

The MLLM-ddPCR algorithm was implemented as a standalone software tool using Python and packaged with a user-friendly GUI. The GUI allows users to input ddPCR images and visualize segmentation results in real time. It has been optimized from single-threaded to multi-threaded for smoother and more efficient operation. Designed for seamless integration with common laboratory fluorescence microscopes, the software facilitates easy adoption in various experimental setups.

Additionally, the interactive personalized setting is also available for users to adjust settings, We set a config file for GUI code, which is convenient for users to configure the expected λ value, MLLM api, MLLM model and MLLM api-key.

Features of the GUI include:

(1) Real-time visualization: displays segmentation results and classification masks.
(2) Data saving: options to save raw images, background-subtracted images, plot results, and calculated values.
(3) Droplet analysis: continuously displays droplet size and calculated template concentration.
(4) Offline mode: capable of analyzing pre-saved image datasets by reading folders.

### 4.6 Statistical analysis

Statistical analysis was performed using GraphPad Prism software (GraphPad Software). All data are presented as mean ± standard deviation (SD) with n ≥ 3. Hypothesis testing was conducted using a t-test, and significance was defined as p ≤ 0.05.

The $r^2$ value in **Figure 6** represents the coefficient of determination. It was computed using the standard formula for linear regression:

$$r^2 = 1 - \frac{\sum(y_i - \hat{y}_i)^2}{\sum(y_i - \bar{y}_i)^2} \tag{10}$$

where

$y_i$ is the observed value, $\hat{y}_i$ is the predicted value from the regression model, and

$\bar{y}_i$ is the mean of the observed data.

This formula quantifies the proportion of the variance in the dependent variable that is explained by the independent variable(s) in the model.

**Acknowledgments**

The authors are grateful for the funding support from the Science, Technology and Innovation Commission (STIC) of Shenzhen Municipality (SGDX20220530111005039), the National Natural Science Foundation of China (Grants No.62204140), Hong Kong Research Grants Council (project reference: GRF142041824, GRF14216222, GRF14203821). The authors would like to thank BioRender.com for providing the tools used to create the illustrations in this article.

The authors would like to acknowledge Mr. Shanhang Luo (Department of Biomedical Engineering, National University of Singapore), Dr. Jianing Qiu (Department of Biomedical Engineering, The Chinese University of Hong Kong), Dr. Ronjie Zhao and Dr. Meng Yan (State Key Laboratory of Marine Pollution, City University of Hong Kong, Hong Kong, China) for their support in the project development.

Contributions

Y. Wei, Y. Wu, W. Yuan, and M. Xu contributed to the study's conception and design. Y. Wu and Y. Wei developed the MLLM-ddPCR algorithm. F. Qu and Y. Wei conducted the biological experiments. Y. Wei and Y. Wu prepared the datasets. Y. Wu and Y. Wei performed data analysis. Y. Wei and Y. Wu prepared figures and tables. F. Qu and Y. Ho supplied the microfluidic chip and customized microfluidic platform. Y. Wei and Y. Wu wrote the manuscript with contributions from all authors. M. Xu, W. Yuan, and Y. Wei conceived the project and supervised the research.

Corresponding author

Correspondence to Mingkun Xu, Wu Yuan, and Yuanyuan Wei.

**Ethics declarations**

No conflict of interest

**Additional information**

# Supporting Information

# Interpretable Droplet Digital PCR Assay for Trustworthy Molecular Diagnostics

| **Supplementary Table 1** | Runtime calculation record of MLLM-ddPCR |
|---|---|
| **Supplementary Table 2** | Supplement table for Figure S3: Performance comparison of MLLM-ddPCR, SAM-dPCR and Dee-qGFP across ddPCR images |
| **Supplementary Table 3** | Supplement table for Figure 4d: Performance comparison of MLLM-ddPCR, SAM-dPCR and Deep-qGFP across different droplet number levels |
| **Supplementary Table 4** | Supplement table for Figure 4e: Performance comparison of MLLM-ddPCR, SAM-dPCR and Deep-qGFP across different SNR levels |
| **Supplementary Table 5** | Supplement table for Figure 4f: fluorescence-intensity-responsive classifier accuracy calculation |
| **Supplementary Table 6** | MLLM-ddPCR accuracy calculation |
| **Supplementary Table 7** | Comparison of MLLM-ddPCR model and manual counting results for different barcods |

| **Supplementary Figure 1** | MLLM-ddPCR GUI enables real-time and high-throughput analysis of droplets |
|---|---|
| **Supplementary Figure 2** | MLLM-ddPCR generates troubleshooting advice |
| **Supplementary Figure 3** | The images and detection results corresponding to the LoD upper and lower limits. |
| **Supplementary Figure 4** | The influence of dust and air bubbles on the segment process of droplets. |
| **Supplementary Figure 5** | Optimization of input size for dPCR image detection |
| **Supplementary Figure 6** | Performance evaluation of MLLM-ddPCR compared to SAM-dPCR and the fully supervised Deep-qGFP model |
| **Supplementary Figure 7** | The actual performance of GUI for figure 5a2 |
| **Supplementary Figure 8** | The actual performance of GUI for figure 5a2 |
| **Supplementary Figure 9** | The actual performance of GUI for figure 5b1 |
| **Supplementary Figure 10** | The actual performance of GUI for figure 5b2 |
| **Supplementary Figure 11** | Performance of MLLM-ddPCR under varying sample concentrations. |

**Supplementary Table 1. Runtime calculation record of MLLM-ddPCR.**

| Number | Capturing/s | Reading/s | Segmenting/s | Classifying/s | Plotting/s | Saving/s | Total/s |
|---|---|---|---|---|---|---|---|
| 1 | 1 | 0.0055 | 4.1209 | 0.8091 | 1.4346 | 0.7192 | 8.089 |
| 2 | 1 | 0.0040 | 4.2936 | 0.2541 | 0.7875 | 0.8405 | 7.180 |
| 3 | 1 | 0.0060 | 4.3306 | 0.2619 | 0.7179 | 0.6538 | 6.970 |
| 4 | 1 | 0.0043 | 4.2908 | 0.2548 | 0.7154 | 0.6535 | 6.919 |
| 5 | 1 | 0.0045 | 4.5423 | 0.3056 | 0.7694 | 0.7422 | 7.364 |
| 6 | 1 | 0.0177 | 4.6588 | 0.2614 | 0.7629 | 0.6678 | 7.369 |
| 7 | 1 | 0.0048 | 4.6154 | 0.2803 | 0.7311 | 0.6493 | 7.281 |
| 8 | 1 | 0.0060 | 4.6537 | 0.3033 | 0.7159 | 0.7449 | 7.424 |
| 9 | 1 | 0.0041 | 4.8401 | 0.2923 | 0.9475 | 0.6210 | 7.705 |
| 10 | 1 | 0.0050 | 4.7787 | 0.2886 | 0.7121 | 0.6536 | 7.438 |
| Average | 1 | 0.0062 | 4.5125 | 0.3311 | 0.8294 | 0.6946 | 7.374 |

**Supplementary Table 2. Supplement table for Figure S5: Performance comparison of MLLM-ddPCR, SAM-dPCR and Deep-qGFP across ddPCR images.**

| Input image | | Figure1 | Figure2 | Figure3 | Figure4 | Figure5 | Figure6 | Figure7 |
|---|---|---|---|---|---|---|---|---|
| MLLM-ddPCR | Positive | 43 | 10 | 44 | 78 | 49 | 22 | 28 |
| | Negative | 647 | 187 | 311 | 630 | 309 | 187 | 183 |
| | Total | 690 | 197 | 355 | 708 | 358 | 209 | 211 |
| SAM-dPCR | Positive | 630 | 25 | 61 | 140 | 101 | 33 | 31 |
| | Negative | 61 | 173 | 296 | 571 | 257 | 176 | 179 |
| | Total | 691 | 198 | 357 | 711 | 358 | 209 | 210 |
| Deep-qGFP | Positive | 44 | 10 | 40 | 90 | 45 | 24 | 27 |
| | Negative | 63 | 2 | 6 | 8 | 4 | 1 | 3 |
| | Total | 107 | 12 | 46 | 98 | 49 | 25 | 30 |
| Ground truth | Positive | 47 | 10 | 43 | 76 | 47 | 23 | 26 |
| | Negative | 650 | 186 | 309 | 636 | 314 | 184 | 182 |
| | Total | 697 | 196 | 352 | 712 | 361 | 207 | 208 |

**Supplementary Table 3. Supplement table for Figure 6d: Performance comparison of MLLM-ddPCR, SAM-dPCR and Deep-qGFP across different droplet number levels.**

| Input image | | Figure 1 | Figure2 | Figure3 | Figure4 |
|---|---|---|---|---|---|
| Droplet number | | <50 | 50~100 | 100~200 | 200~400 |
| MLLM-ddPCR | Positive | 3 | 6 | 18 | 44 |
| | Negative | 31 | 67 | 174 | 307 |
| | Total | 34 | 73 | 192 | 351 |
| SAM-dPCR | Positive | 19 | 38 | 32 | 55 |
| | Negative | 17 | 35 | 161 | 298 |
| | Total | 36 | 73 | 193 | 353 |
| Deep-qGFP | Positive | 3 | 6 | 19 | 43 |
| | Negative | 1 | 0 | 2 | 6 |
| | Total | 4 | 6 | 21 | 49 |
| Groundtruth | Positive | 3 | 6 | 20 | 44 |
| | Negative | 30 | 70 | 175 | 307 |
| | Total | 33 | 76 | 195 | 351 |

**Supplementary Table 4. Supplement table for Figure 6e: Performance comparison of MLLM-ddPCR, SAM-dPCR and Deep-qGFP across different SNR levels.**

| Input image | | Figure 1 | Figure2 | Figure3 | Figure4 |
|---|---|---|---|---|---|
| SNR (dB) | | 4.035 | 1.090 | -0.521 | -1.779 |
| MLLM-ddPCR | Positive | 50 | 49 | 45 | 43 |
| | Negative | 311 | 315 | 323 | 319 |
| | Total | 361 | 364 | 368 | 362 |
| SAM-dPCR | Positive | 90 | 76 | 63 | 48 |
| | Negative | 275 | 287 | 300 | 313 |
| | Total | 365 | 363 | 363 | 361 |
| Deep-qGFP | Positive | 53 | 52 | 53 | 50 |
| | Negative | 6 | 6 | 7 | 7 |
| | Total | 59 | 58 | 60 | 57 |
| Groundtruth | Positive | 50 | | | |
| | Negative | 311 | | | |
| | Total | 361 | | | |

**Supplementary Table 5. Supplement table for Figure 4f: fluorescence-intensity-responsive classifier accuracy calculation.**

|  | Groundtruth positve | Groundtruth negative |
|---|---|---|
| Predicted positive | 296 | 9 |
| Predicted negative | 6 | 293 |

$$ACC = \frac{TP + TN}{TP + TN + FP + FN} = 97.52\%$$

**Supplementary Table 6. MLLM-ddPCR accuracy calculation.**

| Input image | TP | TN | FP | FN |
|---|---|---|---|---|
| 1 | 43 | 640 | 0 | 7 |
| 2 | 10 | 186 | 0 | 1 |
| 3 | 43 | 308 | 1 | 3 |
| 4 | 76 | 628 | 2 | 2 |
| 5 | 47 | 309 | 2 | 0 |
| 6 | 22 | 185 | 0 | 2 |
| 7 | 25 | 180 | 3 | 3 |

$$ACC = \frac{TP + TN}{TP + TN + FP + FN} = 99.05\%$$

**Supplementary Table 7. Comparison of MLLM-ddPCR model and manual counting results for different barcodes.**

| Barcode | Image | SAM-dPCR model | | | | Ground truth (manual counting) | | Sum |
|---|---|---|---|---|---|---|---|---|
|  |  | Positive count | Positive count error | Negative count | Negative count error | Positive count | Negative count |  |
| 1 | 1 | 24 | 0 | 284 | 2 | 24 | 282 | 306 |
|  | 2 | 10 | 0 | 187 | 1 | 10 | 186 | 196 |
|  | 3 | 16 | 0 | 185 | 1 | 16 | 186 | 202 |
| 2 | 1 | 17 | 2 | 175 | 0 | 17 | 175 | 192 |
|  | 2 | 17 | 0 | 176 | 1 | 17 | 175 | 192 |
|  | 3 | 18 | 2 | 174 | 0 | 20 | 174 | 194 |
| 3 | 1 | 22 | 0 | 187 | 1 | 22 | 186 | 208 |
|  | 2 | 27 | 1 | 184 | 2 | 26 | 182 | 208 |
|  | 3 | 44 | 1 | 311 | 3 | 43 | 308 | 351 |

# Supplementary Figures

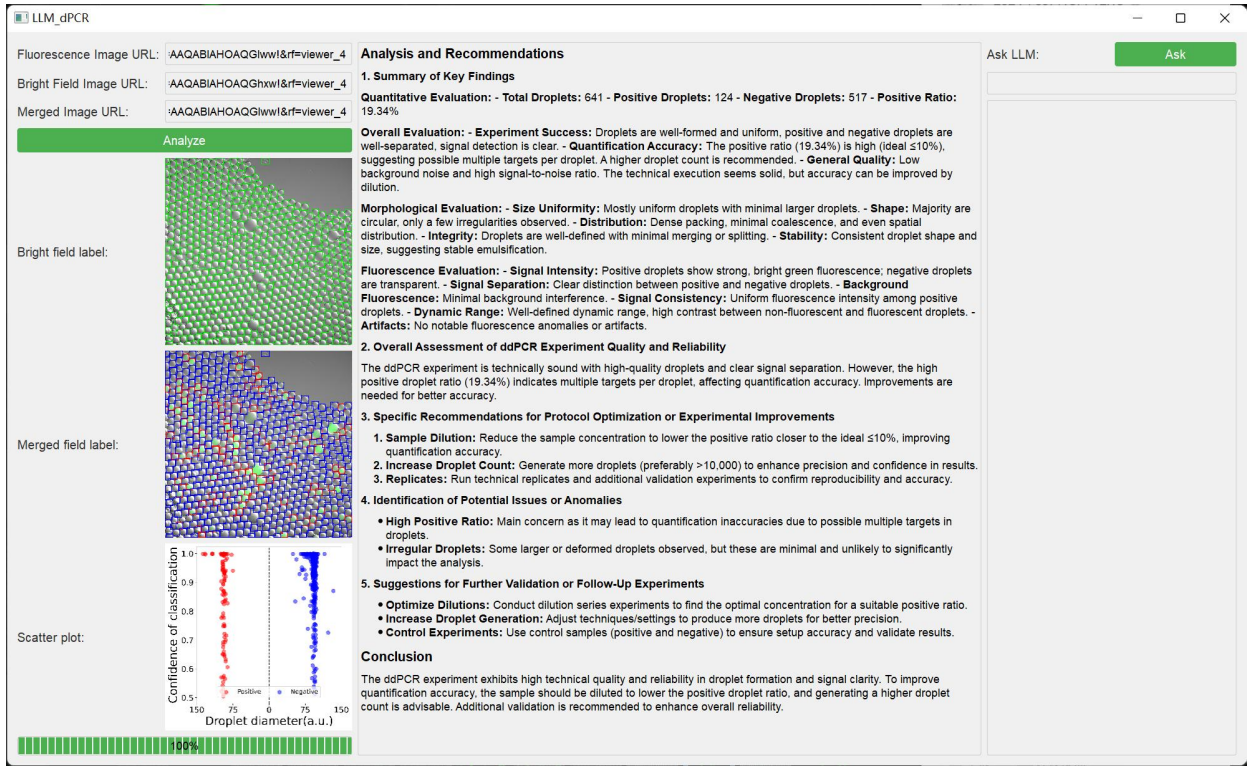

**Figure S1. MLLM-ddPCR GUI enables real-time and high-throughput analysis of droplets.** The GUI displays captured images and analysis results simultaneously. Using the three images uploaded by the user, the GUI performs analyses and presents the current results based on the input images. MLLM analysis outcomes are generated following the completion of image analysis. Furthermore, the GUI allows users to inquire about the results and address any existing questions.

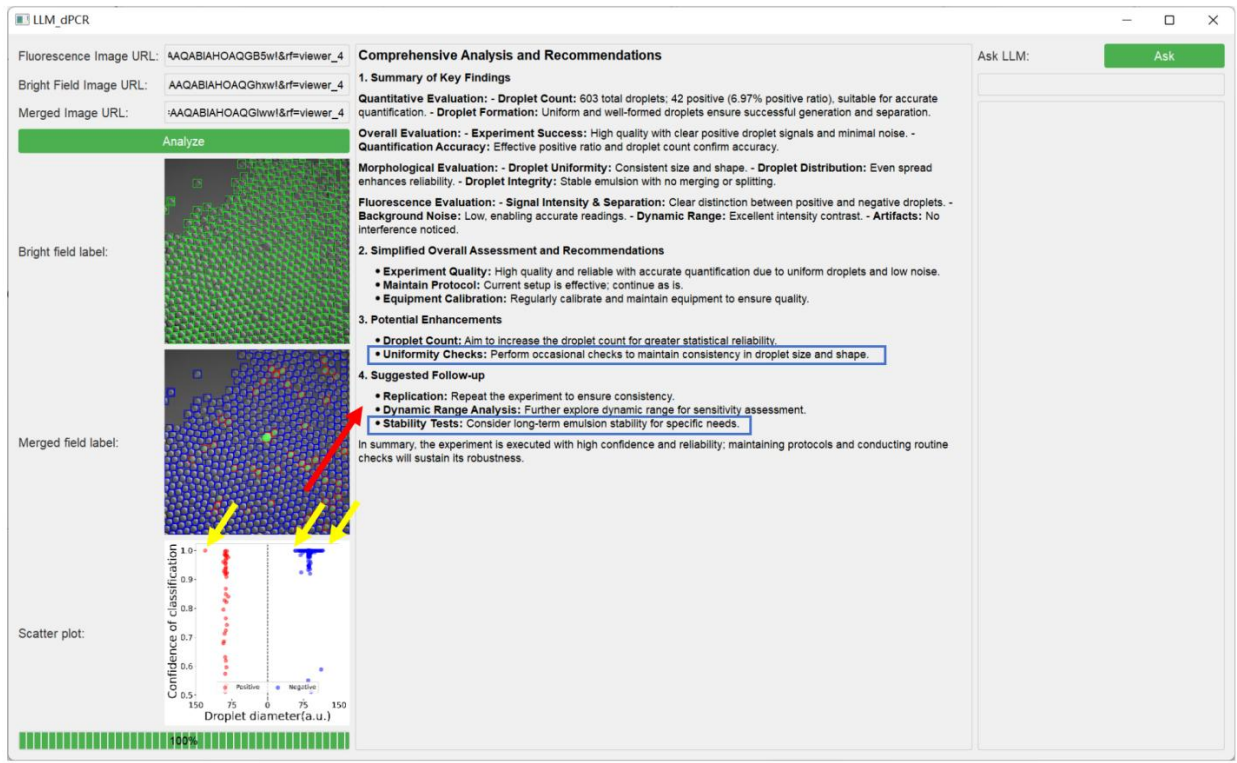

**Figure S2. MLLM-ddPCR generates troubleshooting advice.** Through the experiment analysis, we observe that the droplet diameter is unstable on the scatter diagram. MLLM identified this issue in the observed image and provided potential enhancements and suggestions for follow-up.

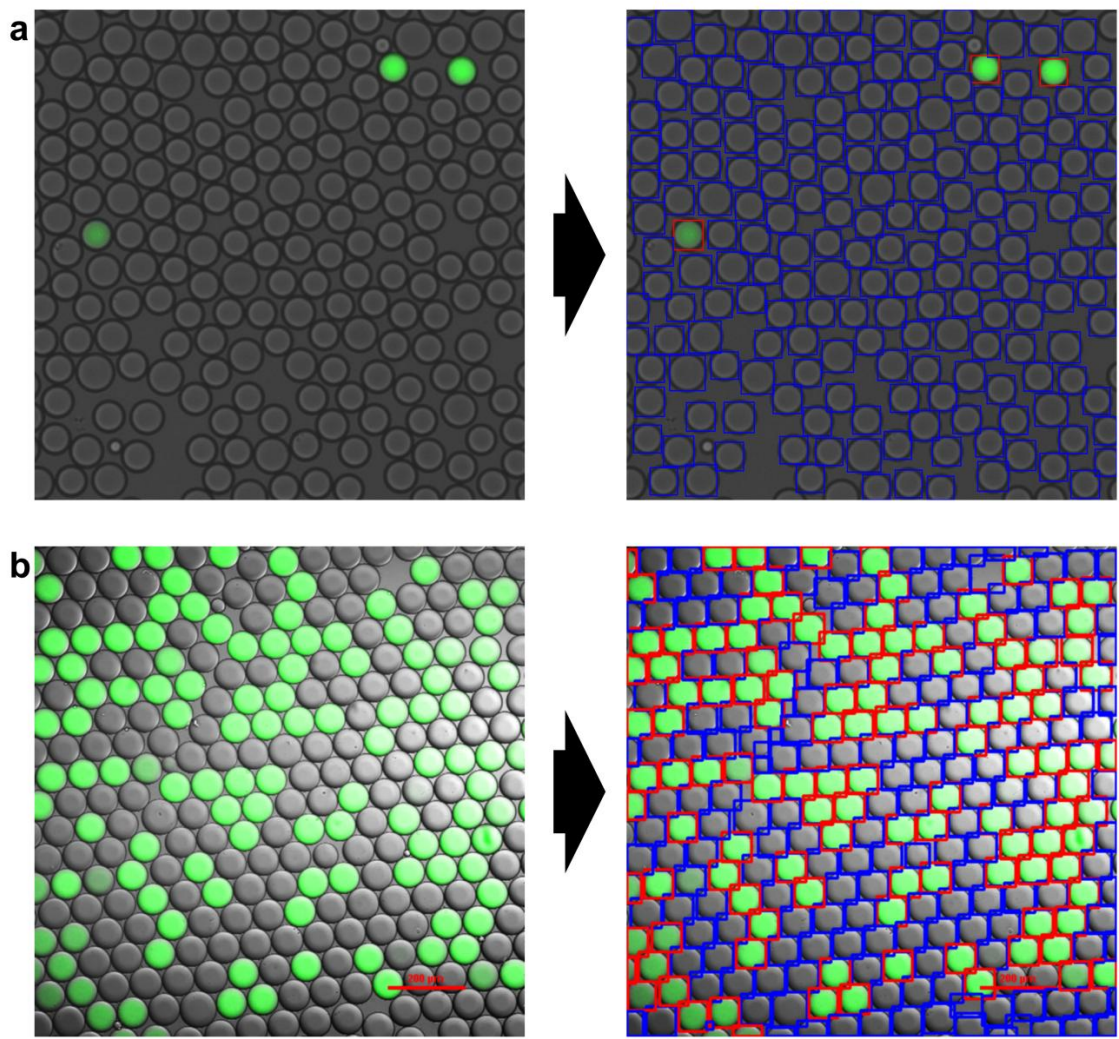

**Figure S3. The images and detection results corresponding to the LoD upper and lower limits.**

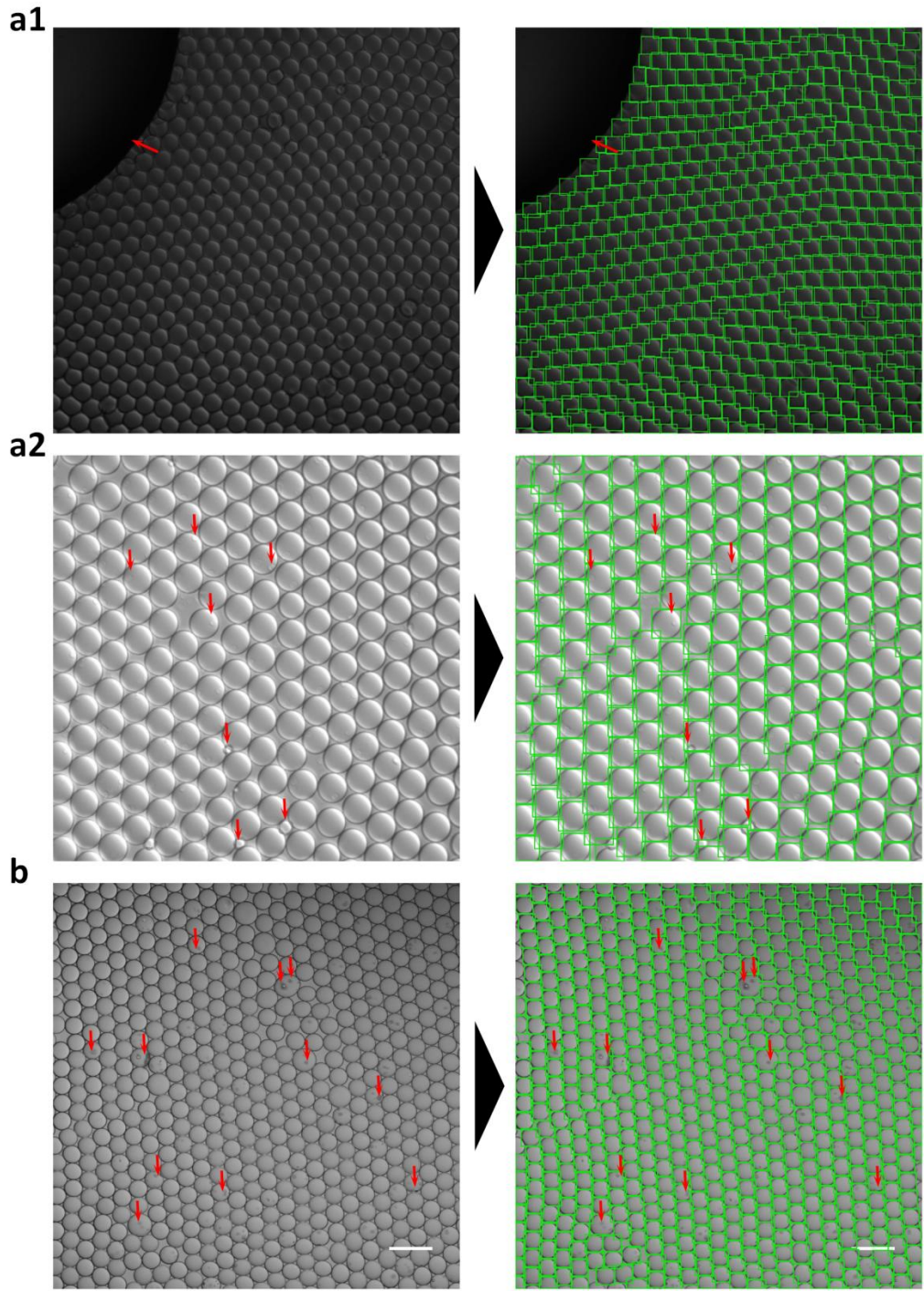

**Figure S4. The influence of dust and air bubbles on the segment process of droplets.** (a1) (a2) The results of droplet segmentation were not affected by air bubbles, and air bubbles did not misidentify the droplet. (b) The results of droplet segmentation were not affected by dust, and dust did not misidentify the droplet.

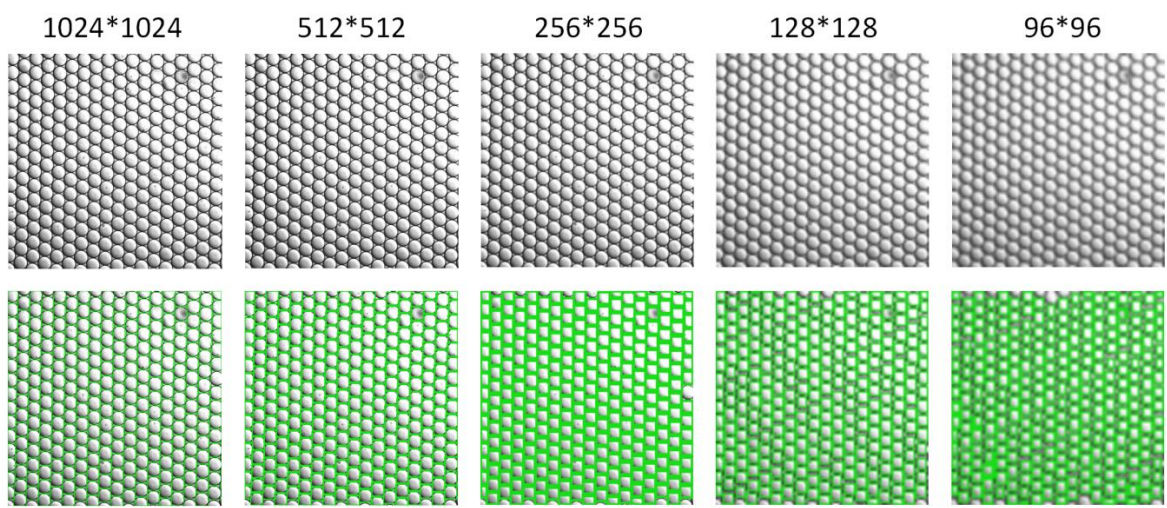

**Figure S5. Optimization of input size for dPCR image detection.** This figure illustrates the impact of varying initial output sizes on the detection performance of our software. We evaluated sizes ranging from 1024×1024 to 64×64 pixels. Results indicate that sizes below 256×256 pixels compromise feature recognition due to small target sizes (<16×16 pixels). The optimal initial sizes for our dataset are determined to be 1024×1024 or 512×512 pixels, offering a balance between image detail and computational efficiency.

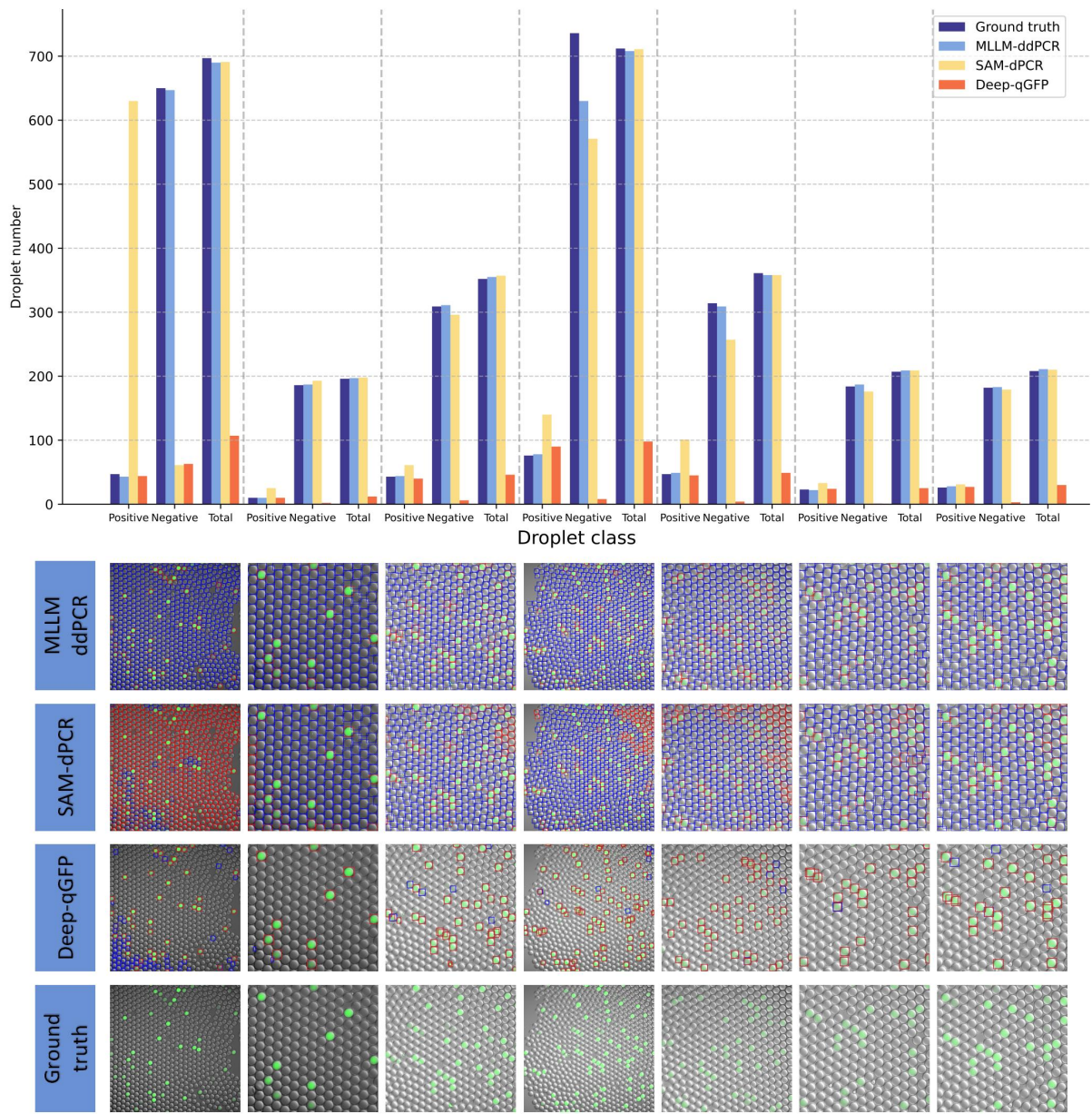

**Figure S6. Performance evaluation of MLLM-ddPCR compared to SAM-dPCR and the fully supervised Deep-qGFP model.** Comparisons were conducted between MLLM-ddPCR, SAM-dPCR, and Deep-qGFP under varying droplet quantities and light conditions. The results indicated that MLLM-ddPCR significantly outperformed SAM-dPCR in classification accuracy. However, Deep-qGFP exhibited a lack of robustness and failed to identify negative droplets.

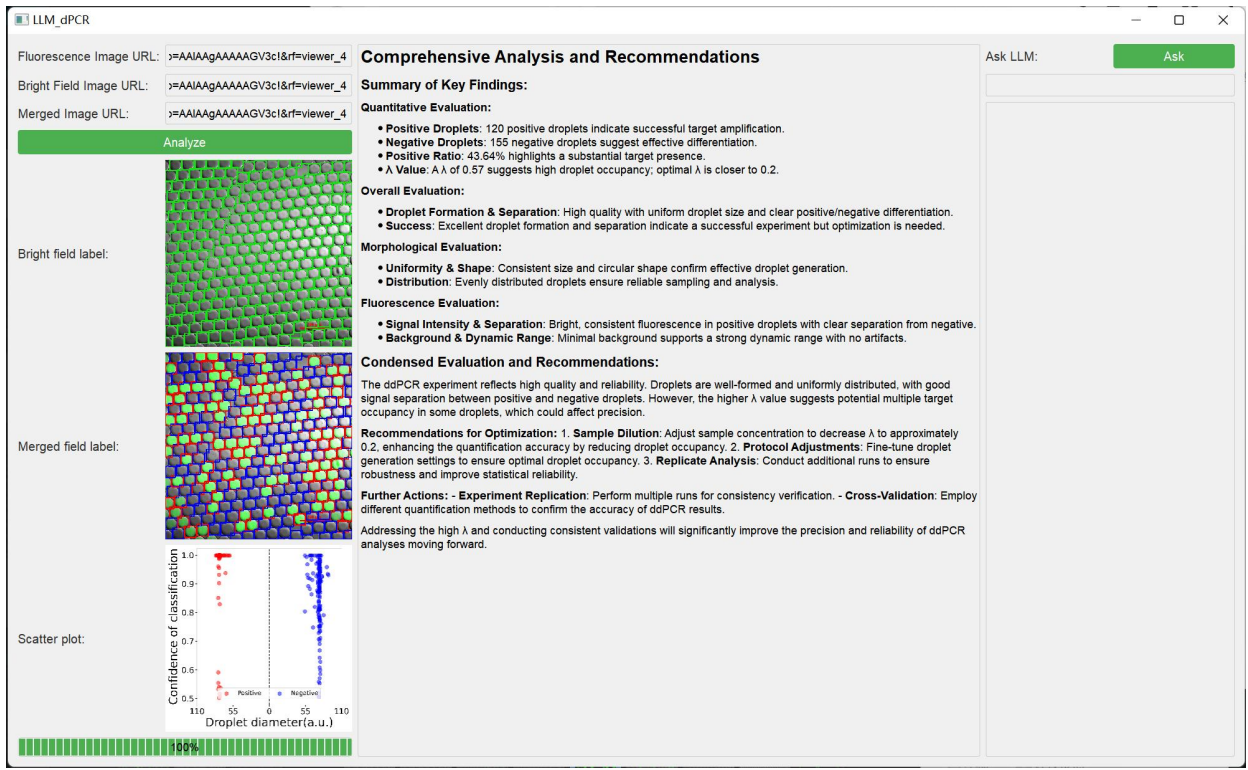

**Figure S7. The actual performance of GUI for figure 5a1.**

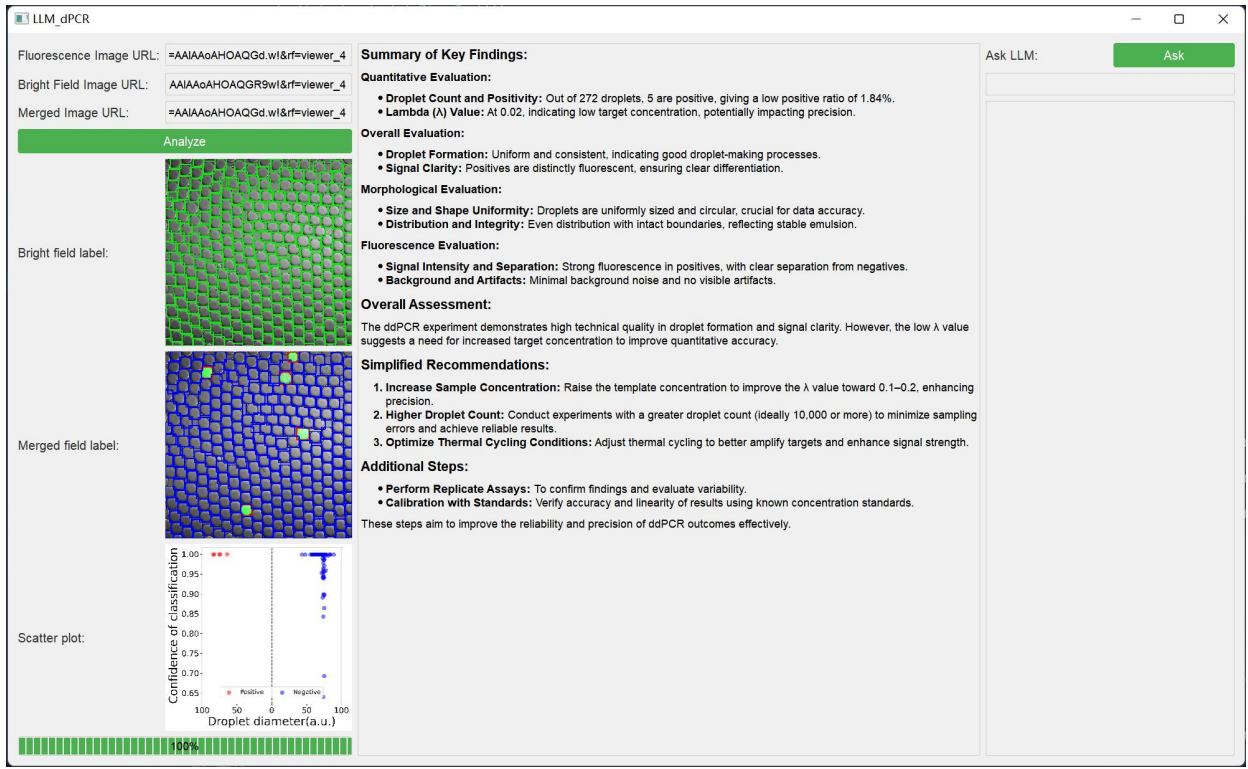

**Figure S8. The actual performance of GUI for figure 5a2.**

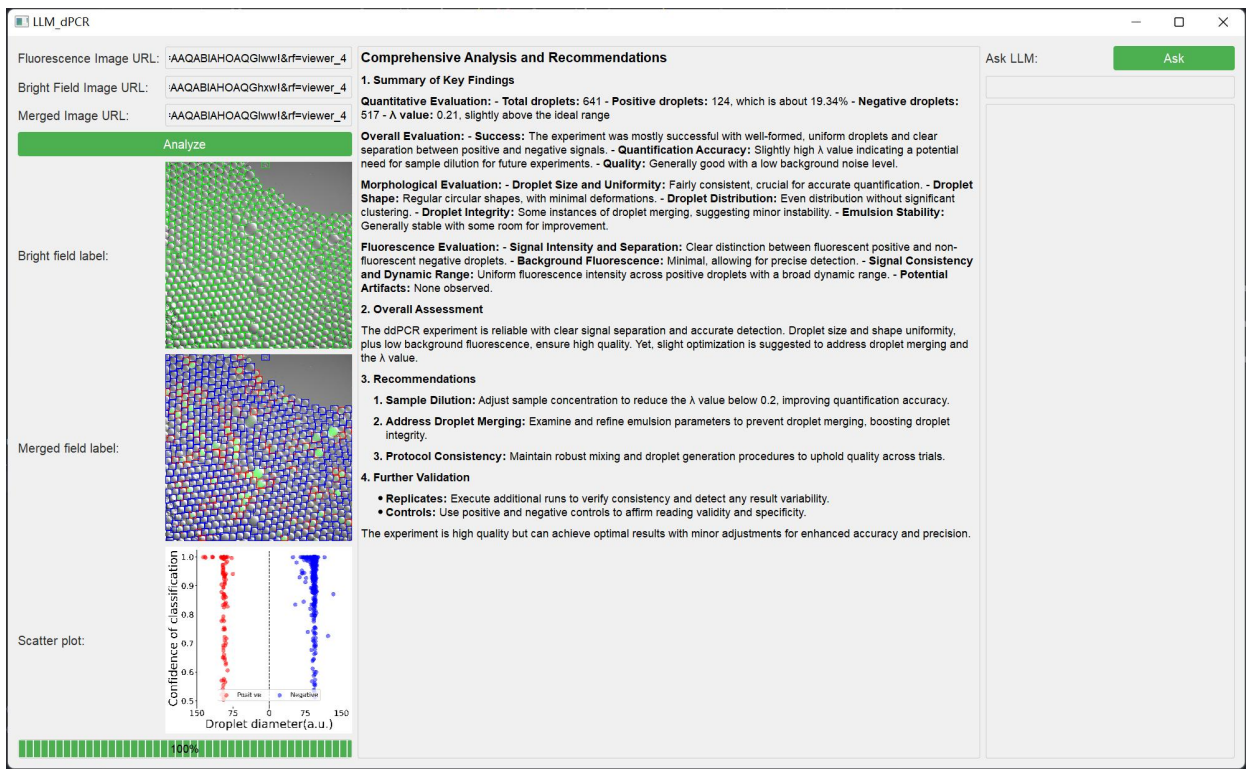

**Figure S9.** The actual performance of GUI for figure 5b1.

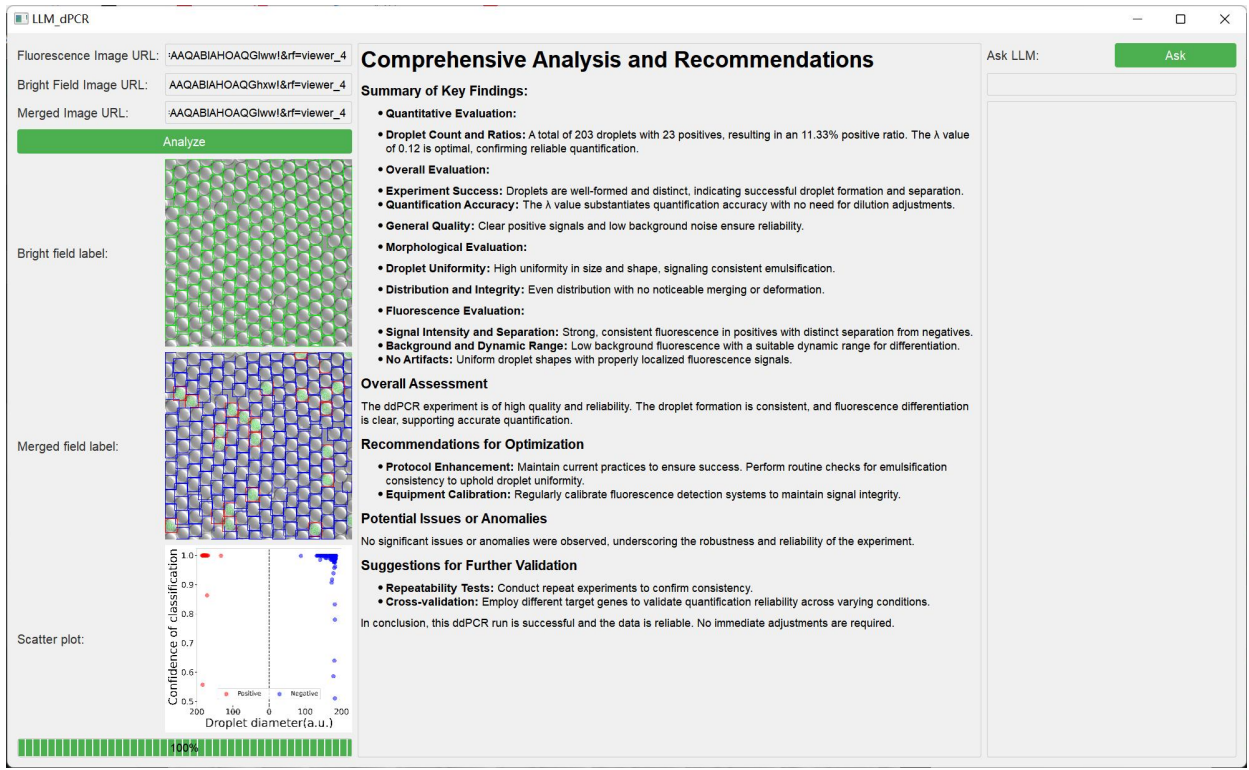

**Figure S10.** The actual performance of GUI for figure 5b2.

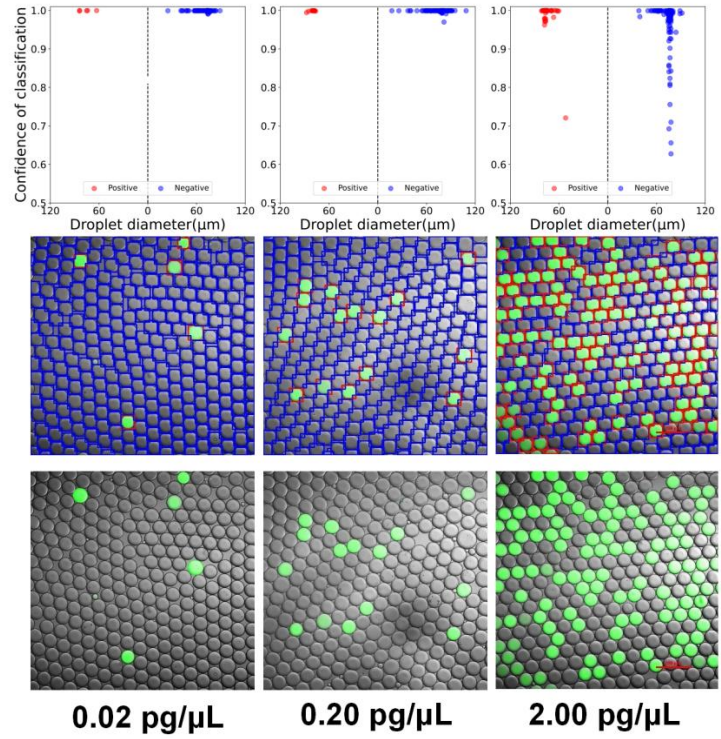

**Figure S11. Performance of MLLM-ddPCR under varying sample concentrations.**

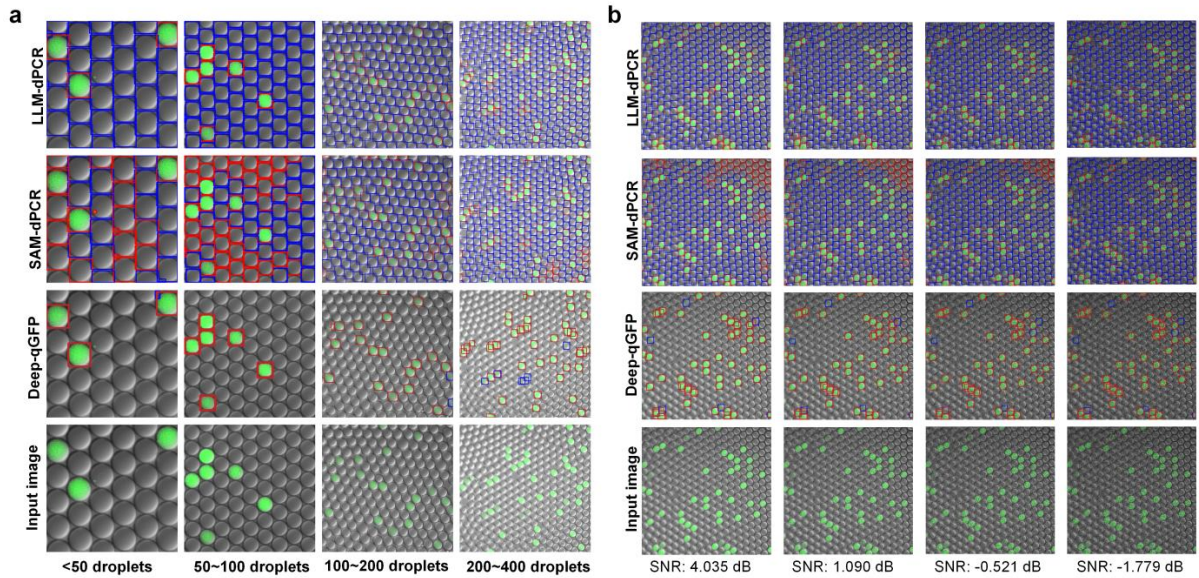

**Figure S12: Supplementary picture for Figure 4.**

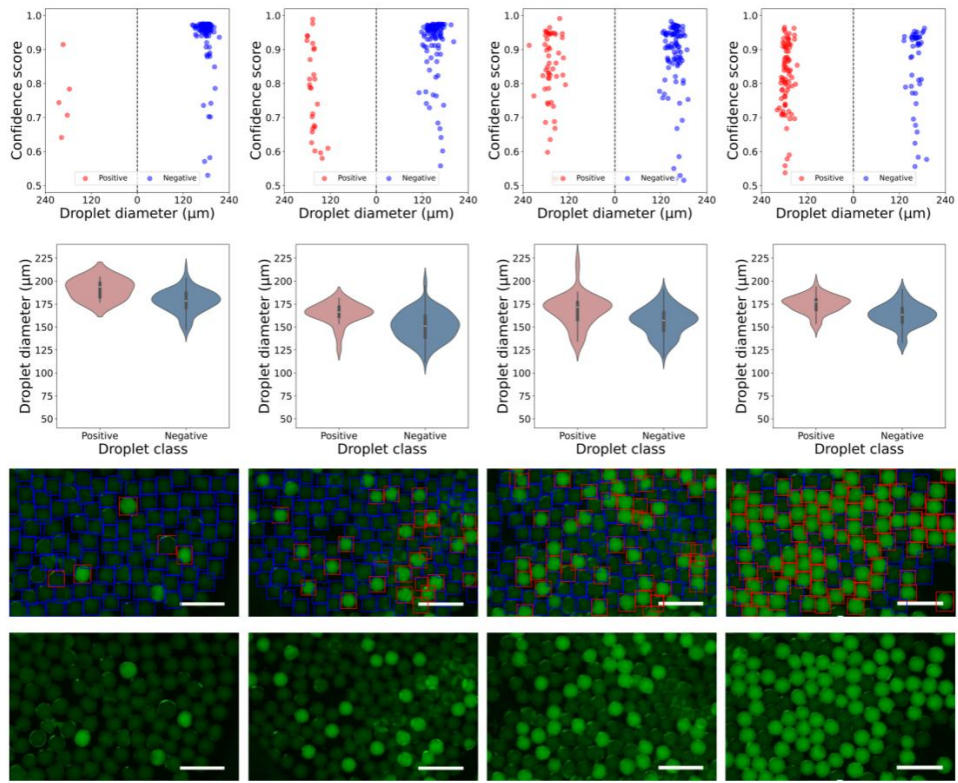

**Figure S13: Output plots for Figure 6a1.**

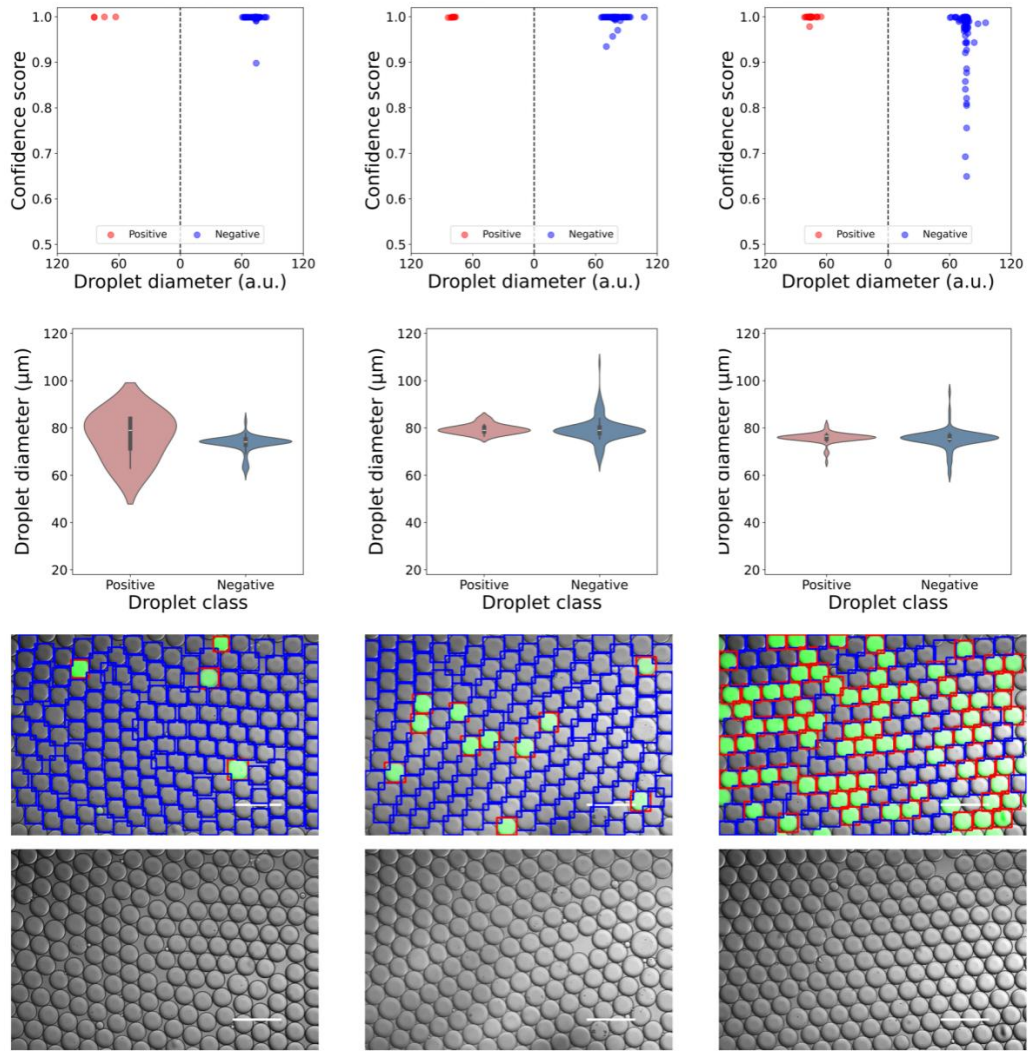

**Figure S14: Output plots for Figure 6b1.**

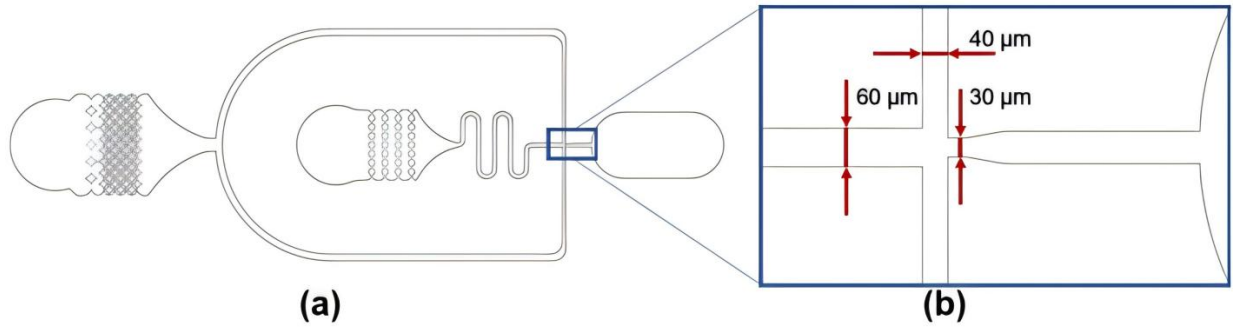

**Fig. S15 Microfluidic chip design for uniform droplet generation and characterization.** (a) The flow-focusing microfluidic chip with a cross-sectional dimension of 30.0 μm in width and 38.5 μm in height was designed and fabricated to generate droplets. (b) Detailed illustration of the throat design in (a) with specific dimensions.